\documentclass[aps,prb,amsmath,amssymb,citeautoscript,reprint]{revtex4-2}
\usepackage[table]{xcolor}
\usepackage{amsmath, amssymb, amsthm}
\usepackage{graphicx, float, xcolor, pgf}
\usepackage{physics, siunitx}
\usepackage{textcomp}
\usepackage[caption=false, justification=centerlast]{subfig}
\let\autocite\cite
\makeatletter
\let\oldtheequation\theequation
\renewcommand\tagform@[1]{\maketag@@@{\ignorespaces#1\unskip\@@italiccorr}}
\renewcommand\theequation{(\oldtheequation)}
\makeatother

\usepackage{hyperref}
\hypersetup{
    pdfauthor={Amartya Bose},
    pdftitle={Study of Exciton Transport in B850 LH2 Ring},
    colorlinks=true,
    allcolors=.
}

\begin{document}
\title{Tensor Network Path Integral Study of Dynamics in B850 LH2 Ring with Atomistically Derived Vibrations}
\author{Amartya Bose}
\email{amartyab@princeton.edu}
\email{amartya.bose@gmail.com}
\affiliation{Department of Chemistry, Princeton University, Princeton, New Jersey 08544}
\author{Peter L. Walters}
\email{peter.l.walters2@gmail.com}
\affiliation{Department of Chemistry, University of California, Berkeley, California 94720}
\affiliation{Miller Institute for Basic Research in Science, University of
  California Berkeley, Berkeley, California 94720}
\thanks{Both authors contributed equally to this work.}
\allowdisplaybreaks

\begin{abstract}

    The recently introduced multisite tensor network path integral (MS-TNPI)
    allows simulation of extended quantum systems coupled to dissipative
    media. We use MS-TNPI to simulate the exciton transport and the absorption
    spectrum of a B850 bacteriochlorophyll (BChl) ring. The MS-TNPI network is
    extended to account for the ring topology of the B850 system. Accurate
    molecular dynamics-based description of the molecular vibrations and the
    protein scaffold is incorporated through the framework of Feynman-Vernon
    influence functional. To relate the present work with the excitonic
    picture, an exploration of the absorption spectrum is done by simulating it
    using approximate and topologically consistent transition dipole moment
    vectors. Comparison of these numerically exact MS-TNPI absorption spectra
    are shown with second-order cumulant approximations. The effect of
    temperature on both the exact and the approximate spectra is also explored.

\end{abstract}
\maketitle

\section{Introduction}\label{sec:intro}

Photosynthesis in plants, bacteria and algae involves light harvesting
complexes. Solar energy creates excitons in these so-called ``antenna
complexes,'' which are subsequently transported to the reaction center.
Understanding this transport process involves the study of the couplings of the
molecular vibrations and the protein scaffold that holds the complex together
and their impact on the dynamics. Approximate simulations of these excitation
energy transfer (EET) processes are often performed using the
Redfield~\cite{ishizakiAdequacyRedfieldEquation2009} and F\"orster resonance
energy transfer (FRET)~\cite{novoderezhkinTheTheoryofForster1996}. While decent
in certain parameter regimes, their accuracy cannot be implicitly assumed. The
complexity of simulating quantum dynamics accurately grows exponentially with
the number of dimensions. For EET systems, the Hilbert space has a large
dimensionality, which consequently makes simulating these systems
computationally challenging.

Typically, rigorous wave function-based methods like the density matrix
renormalization group (DMRG)~\cite{whiteDensityMatrixFormulation1992,
whiteRealTimeEvolutionUsing2004,
schollwockDensitymatrixRenormalizationGroup2005,
schollwockDensitymatrixRenormalizationGroup2011a,
jiangFiniteTemperatureDynamical2020}, the multiconfiguration time-dependent
Hartree (MCTDH)~\cite{beckMulticonfigurationTimedependentHartree2000} and its
multi-layer extension
(ML-MCTDH)~\cite{wangMultilayerFormulationMulticonfiguration2003} have often
been used to simulate the dynamics of extended systems. These methods decompose
the system using various tensor networks to provide a compressed
representation. Though they have been used to study the systems in presence of
vibrational manifolds~\cite{renTimeDependentDensityMatrix2018}, the cost
increases with the number of such vibrational modes and the temperature of the
simulation. This is because the wave function-based approaches proceed by
truncating the basis set corresponding to the bath. In such a framework,
incorporating a continuum of such states at a finite temperature becomes
computationally very challenging~\cite{tanimuraNumericallyExactApproach2020}.

Quantum systems coupled to vibrational dissipative manifolds are most often
simulated using reduced density matrix methods. Foremost among these are the
hierarchical equations of motion (HEOM)~\cite{tanimuraTimeEvolutionQuantum1989,
tanimuraNumericallyExactApproach2020,
yanEfficientPropagationoftheHierarchical2021} and quasi-adiabatic propagator
path integral (QuAPI)~\cite{makriTensorPropagatorIterative1995,
makriTensorPropagatorIterative1995a}. Though historically, HEOM has been
exceptionally popular for simulating large quantum
systems~\cite{strumpferLightHarvestingComplexII2009,
strumpferEffectCorrelatedBath2011, strumpferExcitedStateDynamics2012}, recent
work on path integrals~\cite{makriModularPathIntegral2018,
kunduRealTimePathIntegral2020, kunduOriginVibrationalFeatures2021,
boseMultisiteDecompositionTensor2022} has made it possible to study these
systems as well. Notably, the modular path integral
(MPI)~\cite{makriModularPathIntegral2018} has been used to study the exciton
transfer in bacteriochlorophyll aggregates~\cite{kunduRealTimePathIntegral2020}.
The semiclassical partially linearized density matrix path integral approach
along with accurate spectral density have also been used by
\citeauthor{leeSemiclassicalPathIntegral2016}~\cite{leeSemiclassicalPathIntegral2016}
to study the Fenna-Matthew-Olson complex.

We have recently developed a multisite tensor network path integral method
(MS-TNPI)~\cite{boseMultisiteDecompositionTensor2022} using the framework of
tensor network path
integral~\cite{boseTensorNetworkRepresentation2021,
bosePairwiseConnectedTensor2022, strathearnEfficientNonMarkovianQuantum2018,
jorgensenExploitingCausalTensor2019}. MS-TNPI starts with a tensor network
decomposition of the system, similar to what is commonly used in
DMRG~\cite{schollwockDensitymatrixRenormalizationGroup2005,
schollwockDensitymatrixRenormalizationGroup2011,
schollwockDensitymatrixRenormalizationGroup2011a,
paeckelTimeevolutionMethodsMatrixproduct2019} and extends it to incorporate the
Feynman-Vernon influence functional~\cite{feynmanTheoryGeneralQuantum1963}. In
order to achieve this, the decomposition along the system spatial dimension is
extended to the temporal dimension creating a 2D tensor network. It is along
this temporal dimension that the influence functional is applied in the form of
a matrix product operator (MPO). MS-TNPI, being based on the Feynman-Vernon
influence functional, can handle arbitrary spectral densities describing the
dissipative environment. The resultant 2D MS-TNPI network can be efficiently
contracted to yield the time-dependent reduced density matrix corresponding to the
extended quantum system represented in the form of a matrix product state
(MPS). This representation contains the full Hilbert space of the system. Thus,
the method is not limited to problems that can only be formulated in the first
excitation Frenkel subspace. It allows for simulations of higher order spectra
and many-body observables while ensuring that the dissipative medium is still
treated in a numerically exact manner.

To accurately simulate the excitonic dynamics in B850, high quality
parameterizations of the environment are essential. Much effort has gone into
such studies. Starting from simulations of photosynthetic complexes using model
spectral densities~\cite{ishizakiTheoreticalExaminationQuantum2009,
ishizakiAdequacyRedfieldEquation2009}, studies have incorporated descriptions
using experiments like fluorescence line narrowing
spectra~\cite{ratsepElectronPhononVibronic2007,
boseAllModeQuantumClassical2020} and fully theoretically simulated spectral
densities~\cite{olbrichTimeDependentAtomisticView2010,
olbrichTheorySimulationEnvironmental2011, maityDFTBMMMolecular2020}. The
benefit of using a theoretically simulated spectral density is the internally
consistent of treatment the high frequency ``quantum'' region comprising of
rigid vibrations and low frequency ``classical'' region primarily made of
ro-translational modes. Both regions need to be accounted for to obtain
accurate dynamics. In this paper, we study the dynamics and
absorption spectrum of the exciton transport in the B850 ring of LH2. Molecular
dynamics-based descriptions of the dissipative medium coupled to the
chlorophyll ring are available in the form of spectral
densities~\cite{olbrichTimeDependentAtomisticView2010}. This spectral density
captures the effect of the high frequency rigid molecular vibrations as well as
the ro-translational modes primarily coming from the protein scaffold using
molecular dynamics.

This paper is organized as follows. Section~\ref{sec:method} summarizes the
MS-TNPI method and network that is used for the simulations. We describe how the
ring conformations can be included after making minor modifications to the
propagator. We also show how the absorption spectrum can be calculated using
MS-TNPI. This tensor network formulation, utilizing the many-body reduced
density matrix corresponding to the extended system, allows a single simulation
to give the entire spectrum. Then we discuss the B850 system under study and
the simulation results in Sec.~\ref{sec:results}. We analyse the symmetries
present in the dynamics and demonstrate the spectra corresponding to different
approximations. We also explore the temperature effects on the absorption
spectrum under the assumption that the solvent spectral density invariant over
the temperature range and show how the behavior of the approximate spectra is
qualitatively different. The simulations are computationally quite cheap,
probably owing to the structure of the Frenkel model. We also discuss the
differences between this model and the well-known Ising model in presence of a
dissipative medium. Finally, we end the paper with some concluding remarks in
Sec.~\ref{sec:conclusions}.

\section{Method}\label{sec:method}

\subsection{Multisite Tensor Network Path Integral}\label{sec:PI_Exp}

Consider an extended quantum system consisting of $P$ sites coupled with
vibrational modes described by the following Hamiltonian:
\begin{align}
    \hat{H} & = \hat{H}_0 + \sum_{i=1}^P \hat{V}_i
\end{align}
where $\hat{H}_0$ is the Hamiltonian describing the quantum system and
$\hat{V}_i$ captures the interaction of the $i$\textsuperscript{th} site with
its local vibrational modes.

For an EET process, the individual system sites are chromophores. The
$i$\textsuperscript{th} site can be represented by the two states, a ground
state, $\ket{\phi_i^g}$, and an excited state, $\ket{\phi_i^e}$. The Hamiltonian
corresponding to the quantum system consequently can be expressed as a Frenkel
model:
\begin{align}
    \hat{H}_0 = &\sum_{i=1}^P E_i \dyad{e_i} + \sum_{i=1}^{P-1} J_i \left(\dyad{e_{i+1}}{e_i} + \dyad{e_i}{e_{i+1}} \right)\label{eq:frenkel}\\
    \text{where }\nonumber \\
    &\ket{e_i} = \ket{\phi_i^e}\otimes\prod^\otimes_{j\ne i}\ket{\phi_j^g}.\label{eq:1exciton_state}
\end{align}
Here, $\ket{e_i}$, as expressed by the direct product of the site-local basis
in Eq.~\ref{eq:1exciton_state}, is the single-exciton state with the excitation
localized on the $i$\textsuperscript{th} site. The electronic excitation energy
of the $i$\textsuperscript{th} system is $E_i$. Here we have assumed that only
the nearest neighbor units are coupled through the coupling $J_i$. This
typically represents a chain. However, B850 is a ring, so there is an extra
interaction term between the $i=1$ and $i=P$ sites, $H_{1,P} =
J_P\left(\dyad{e_P}{e_1} + \dyad{e_1}{e_P}\right)$.

Under Gaussian response theory, the effect of the dissipative medium can be
mapped onto a bath of harmonic oscillators:

\begin{align}
    \hat{V}_i & = \sum_{l} \frac{p^2_{i,l}}{2m_{i,l}} + \frac{1}{2}m_{i,l}\omega_{i,l}^2 \left(x_{i,l} - \frac{c_{i,l} \hat{s}_i}{m_{i,l} \omega_{i,l}^2}\right)^2\label{eq:harmonicbath},
\end{align}
where $\omega_{i,l}$ and $c_{i,l}$ are the frequency and coupling of the
$l$\textsuperscript{th} mode of the $i$\textsuperscript{th} site, respectively.
The system operator, $\hat{s}_i$ associated with the $i$\textsuperscript{th}
site, couples the site with its local vibrations. For EETs, the $\hat{s}_i$
operators are typically characterized by $\hat{s}_i\ket{\phi^g_i} = 0$ and
$\hat{s}_i\ket{\phi^e_i} = 1$. The site-vibration interaction is described by a
spectral density~\cite{caldeiraPathIntegralApproach1983,
makriLinearResponseApproximation1999}
\begin{align}
    J_i(\omega) & = \frac{\pi}{2}\sum_l\frac{c_{i,l}^2}{m_{i,l}\omega_{i,l}}\delta(\omega-\omega_{i,l}),
\end{align}
which is related to the energy-gap autocorrelation function obtained using
classical trajectory-based methods.

For the case of the B850 ring, \citet{olbrichTimeDependentAtomisticView2010}
have simulated the correlation function using classical molecular dynamics for
the trajectories and ZINDO/S-CIS for the excitation energy. This correlation
function on the $i$\textsuperscript{th} site was subsequently fit as a sum of
exponentials and damped oscillations:
\begin{align}
    C_i(t) &= \sum_{j=1}^{N_\text{exp}} \eta_j e^{-\gamma_j t} + \sum_{j=1}^{N_\text{osc}} \tilde\eta_j\cos(\omega_j t) e^{-\tilde\gamma_j t}\label{eq:corrfunc}
\end{align}
where $N_\text{exp}$ is the number of exponentials and $N_\text{osc}$ is the
number of damped oscillations needed to fit the correlation function. Here,
$\eta_j$ is the strength of the $j$\textsuperscript{th} exponential with a
decay rate $\gamma_j$. The tildes correspond to the damped oscillations.
The spectral density is given as a cosine transform of this correlation function and has the following
form~\cite{olbrichTimeDependentAtomisticView2010,olbrichTheorySimulationEnvironmental2011}:
\begin{align}
    J_i(\omega) & = \frac{2}{\hbar} \tanh\left(\frac{\hbar\omega\beta}{2}\right) \int_0^\infty C_i(t) \cos(\omega t) \dd{t}\label{eq:spect_base}\\
                & = \frac{2}{\hbar}\tanh\left(\frac{\hbar\omega\beta}{2}\right)\left[\sum_{j=1}^{N_\text{exp}} \frac{\eta_j\gamma_j}{\gamma_j^2 + \omega^2} \right.\nonumber \\
              & \left.+ \sum_{j=1}^{N_\text{osc}} \frac{\tilde{\eta}_j\tilde{\gamma}_j}{2\left(\tilde\gamma_j^2 + (\omega - \omega_j)^2\right)}\right]
\end{align}
The correlation functions and thus the spectral densities can, in general, be
different for the various sites. However, for B850, due to cylindrical symmetry
of the complex, all the spectral densities (and correlation functions) are
identical according to~\citet{olbrichTimeDependentAtomisticView2010}.
Experimentally, these baths are often reported by their Huang-Rhys factors.
Many simulations have been previously done with HEOM using a Drude-Lorentz
spectral density~\cite{strumpferEffectCorrelatedBath2011,
strumpferLightHarvestingComplexII2009}, which is a specialization of the above
form with $N_\text{exp}=1$ and $N_\text{osc}=0$. Of course, such a form is far
less flexible in accounting for the full physics of the problem. In particular,
it misses out on the contributions from the rigid molecular vibrations.

It should be noted that there are a variety of ways of evaluating quantum
correlation functions from purely classical data as summarized
in Refs.~\cite{kimEvaluationQuantumCorrelation2002,
kimEvaluationQuantumCorrelation2006}. It has been recently shown that the
so-called harmonic approximation, where the hyperbolic tangent is replace by its
high temperature limit, yields better agreement with the quantum
correlation function. It has also been shown to better maintain the temperature independence
of the spectral density~\cite{valleauOnTheAlternatives2012}. Additionally, the
approach employed by~\citet{olbrichTimeDependentAtomisticView2010} suffers from
the "geometry mismatch" problem, i.e., the potential surface used for the
classical molecular mechanics (MM) part of the simulation does no have the same normal modes and
frequencies as the corresponding quantum potential. In particular, it has been
shown that for a gas phase BChl molecule the frequencies predicted by the MM
surface are at substantially higher then those corresponding to the quantum surface~\cite{leeModelingElectronicNuclearInteractions2016}.   
Newer methods of calculating spectral densities designed to remedy these
issues are available~\cite{leeModelingElectronicNuclearInteractions2016,
maityDFTBMMMolecular2020}. It is not entirely clear to what extent these deficiencies in the
spectral density effect observables like the population dynamics and spectra. As
MS-TNPI is derived completely independently from the spectral density used, it
offers an excellent means to investigate the effects of these improved spectral
densities on observables of interest. While not done here, this will be the
topic of future research.

The reduced density matrix of the extended quantum system can be
represented as a path integral expression:
\begin{align}
    \tilde\rho(S^\pm_N, N\Delta t) & = \Tr_\text{bath} \mel{S^+_N}{\rho(N\Delta t)}{S^-_N} \\ \label{eq:RDM_Simple}
                                   & = \sum_{S_0^\pm}\sum_{S^\pm_1}\cdots\sum_{S^\pm_N-1} \tilde\rho(S^\pm_0, 0) P_{S_0^\pm\cdots S^\pm_N} \\
                                   & = \sum_{S_0^\pm}\sum_{S^\pm_1}\cdots\sum_{S^\pm_N-1} \tilde\rho(S^\pm_0, 0) P^{(0)}_{S_0^\pm\cdots S^\pm_N} F\left[\left\{S^\pm_n\right\}\right]\label{eq:RDM_Full}
\end{align}
where $\tilde\rho$ is the reduced density matrix at an arbitrary time,
$P_{S^\pm_0\cdots S^\pm_N}$ is the path amplitude tensor,
$P^{(0)}_{S^\pm_0\cdots S^\pm_N}$ is the bare path amplitude tensor and $F$ is
the Feynman-Vernon influence functional~\cite{feynmanTheoryGeneralQuantum1963}.
The system states at the $n$\textsuperscript{th} time point are collectively
denoted by $S_n^\pm$. (The superscript of ``$\pm$'' denotes the combined
forward-backward state, with the ``$+$'' and ``$-$'' coordinates defining the
bra and ket sides of the reduced density matrix respectively. The state of the
$i$\textsuperscript{th} site at the $n$\textsuperscript{th} time point will be
denoted by $s^\pm_{i,n}$. In these notations, the first index will be the
spatial index and the second index will be the temporal one. The
forward-backward state of the $i$\textsuperscript{th} site, $s^\pm_{i,n}$, can
take values corresponding to $\dyad{\phi^g_i}$, $\dyad{\phi^g_i}{\phi^e_i}$,
$\dyad{\phi^e_i}{\phi^g_i}$, or $\dyad{\phi^e_i}$.) The key terms appearing
here are the bare path amplitude tensor, the influence functional and the path
amplitude tensor. The bare path amplitude tensor, $P^{(0)}_{S^\pm_0\cdots
S^\pm_N}$, contains the full information of the system independent of the
solvent. It is given by:
\begin{align}
    P^{(0)}_{S^\pm_0\cdots S^\pm_N} & = K(S^\pm_0, S^\pm_1, \Delta t) K(S^\pm_1, S^\pm_2, \Delta t)\nonumber \\
                                    & \times\cdots K(S^\pm_{N-1}, S^\pm_N, \Delta t),
\end{align}
where $K$ is the so-called ``forward-backward propagator'' obtained from a
direct product of the forward, $U$, and backward, $U^\dag$, system propagators,
\begin{align}
    K(S^\pm_n, S^\pm_{n+1}, \Delta t) &= U(S^+_n, S^+_{n+1}, \Delta t)\,U^\dag(S^-_n, S^-_{n+1}, \Delta t).
\end{align}
The influence functional, $F$, encodes the interaction of the system with the
solvent. Since the vibrational modes are site local, it can be expressed as a
product of site-specific influence functionals:
\begin{align}
    F\left[\left\{S^\pm_n\right\}\right] &= \prod_{i=1}^P F_i\left[\left\{s^{\pm}_{i,n}\right\}\right],
\end{align}

where

\begin{align}
    F_i\left[\left\{s^{\pm}_{i,n}\right\}\right] & = \exp\left(-\frac{1}{\hbar}\sum_{k} \Delta s_{i,k} \sum_{k'\le k}\left(\Re\left(\eta^i_{k,k'}\right)\Delta s_{i,k'} \right.\right.\nonumber \\
                                            & \left. \vphantom{\sum_{k\le k'}}\left. + 2i\Im\left(\eta^i_{k,k'}\right) \bar{s}_{i,k'}\right)\right).
\end{align}
The bath response function, $C_i(t)$ in Eq.~\ref{eq:corrfunc}, discretized
along the QuAPI system
path~\cite{makriTensorPropagatorIterative1995,makriTensorPropagatorIterative1995a}
for the $i$\textsuperscript{th} site is given by $\eta_{kk'}^i$. Additionally, $\Delta
s_{i,k} = s^{+}_{i,k} - s^{-}_{i,k}$ and $\bar{s}_{i,k} =
\tfrac{1}{2}\left(s^{+}_{i,k} + s^{-}_{i,k}\right)$. (The method is, of course,
not tied down to any specific form of the spectral density like the Drude or
the Ohmic forms. The $\eta$-coefficients can be expressed in terms of integrals
over the spectral density.) Because the B850 ring necessitates identical
spectral densities, the $\eta$-coefficients are the same for every site. The
path amplitude tensor, $P_{S^\pm_0\cdots S^\pm_N}$ is effectively the product
of the bare path amplitude tensor and the influence functional. It contains the
full information of the system embedded in the solvent.

\begin{figure}
    \subfloat[MPS representation of the density matrix]{\includegraphics[scale=0.2]{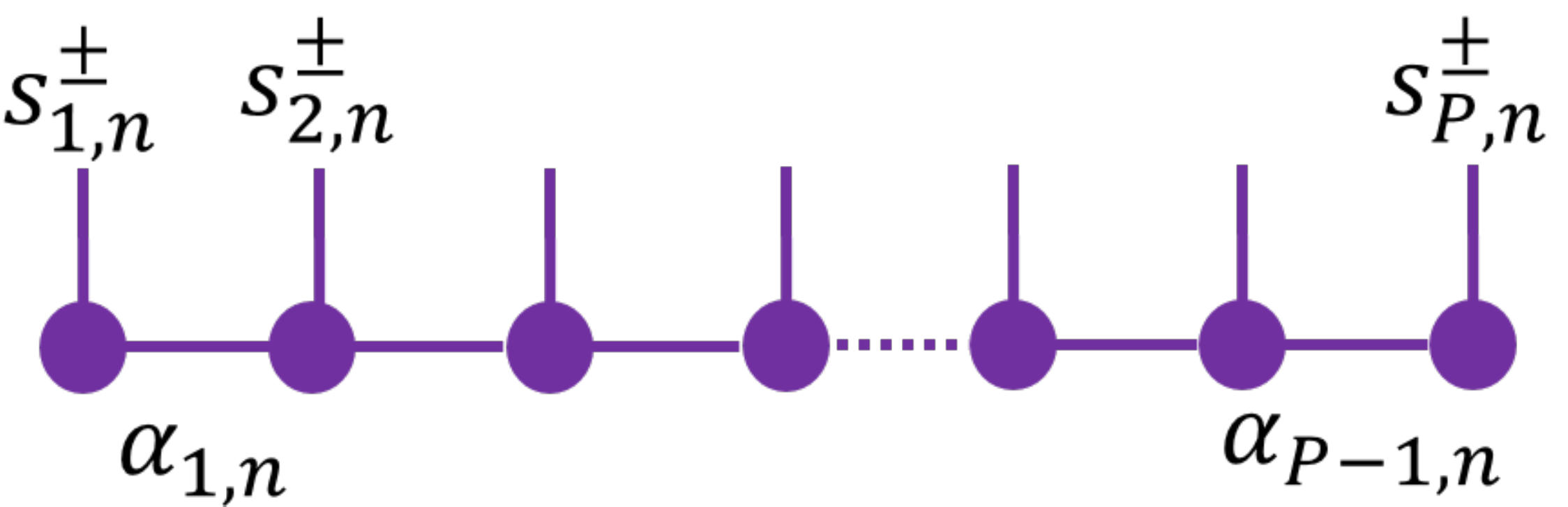}}
    \hspace{2cm}\subfloat[MPO representation of the propagator]{\includegraphics[scale=0.35]{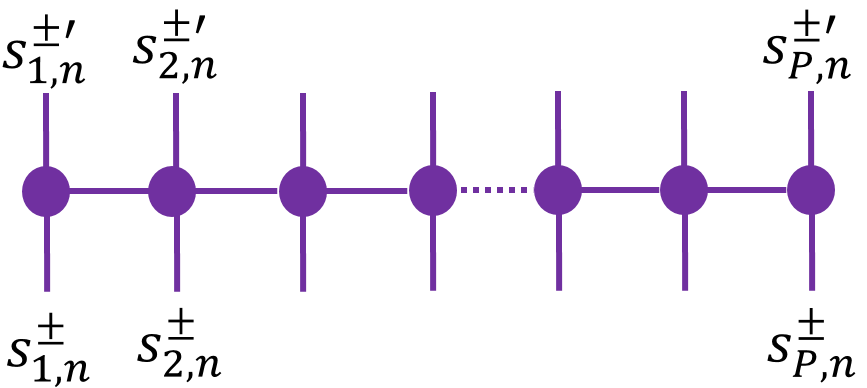}}
    \caption{Matrix product representations of the density matrix and the propagator.}\label{fig:rho_MPS}
\end{figure}

To simplify the discussion, let us briefly neglect the effects of the solvent.
Under this condition, the summations in Eq.~\ref{eq:RDM_Full} can be performed
independently, since the influence functional, which couples the system at
different time points, is omitted when there is no system-solvent interactions.
In this case, the density matrix can be evaluated iteratively, 
\begin{align}
    \tilde\rho(S^\pm_n, n \Delta t)  & = \sum_{S_{n-1}^\pm}\tilde\rho(S^\pm_{n-1}, (n-1)\Delta t) K(S^\pm_{n-1}, S^\pm_{n}, \Delta t). \label{eq:RDM_Iter}
\end{align}
This has the same form as matrix-vector multiplications. The storage and
computational complexity of these expressions grow exponentially with the
number of system sites making direct simulations of extended systems
practically impossible.  However, for many extended systems, the correlations
between system sites decrease rapidly with the distance between them.
Thus, the large tensors (e.g., $\tilde\rho$ and $K$) representing every
particle in the system can be efficiently factored into a network of smaller
tensors corresponding to a single particle each. This fact is widely utilized
by methods like time-dependent
DMRG~\cite{daleyTimedependentDensitymatrixRenormalizationgroup2004} and
time-dependent variational principal
(TDVP)~\cite{haegemanTimeDependentVariationalPrinciple2011,yangTimedependentVariationalPrinciple2020}.
In this representation, the reduced density matrix becomes an MPS, 

\begin{align}
    \tilde{\rho}\left(S_n^\pm, n\Delta t\right) &= \sum_{\left\{\alpha_{(i,n)}\right\}} A_{\alpha_{(1,n)}}^{s^\pm_{1,n}} A_{\alpha_{(1,n)},\alpha_{(2,n)}}^{s^\pm_{2,n}}\cdots A_{\alpha_{(P-1,n)}}^{s^\pm_{P,n}},
\end{align}
the forward-backward propagator an MPO,
\begin{align}
    K\left(S_{n}^\pm, S_{n+1}^\pm, \Delta t\right)  = & \sum_{\left\{\alpha_{(i,n)}\right\}}
    W^{s_{1,n}^\pm, s_{1,n+1}^\pm}_{\alpha_{(1,n)}}  W^{s_{2,n}^\pm,
    s_{2,n+1}^\pm}_{\alpha_{(1,n)},\alpha_{(2,n)} }\nonumber                      
                                 \\
                                                      & \cdots W^{s_{P-1,n}^\pm,
                                     s_{P-1,n+1}^\pm}_{\alpha_{(P-2,n)},\alpha_{(P-1,n)}}W^{s_{P,n}^\pm,
                             s_{P,n+1}^\pm}_{\alpha_{(P-1,n)}},\label{eq:prop_MPO}
\end{align}
and the matrix-vector multiplication becomes an MPO-MPS application. Here,
$\alpha_{i,n}$ is the ``bond'' index that connects the $i$\textsuperscript{th}
site at time-step $n$ to the $(i+1)$\textsuperscript{th} site at the same time
step. It is called a ``spatial'' bond index because it connects points that are
spatially separated. The structures of the MPS and MPO are shown in
Fig.~\ref{fig:rho_MPS}. The maximum and average bond dimension associated with
the $n$\textsuperscript{th} time step is  $m(n)=\max_{i}
\left(\dim(\alpha_{(i,n)})\right)$ and $\bar{m}(n)=\tfrac{1}{P}\sum_i
\dim(\alpha_{(i,n)})$, respectively. The efficiency of these factorizations can
often be characterized by the maximum bond dimension. Roughly speaking, the
smaller the resulting bond dimension the more efficient the MPO/MPS
factorization. 

For most problems, there are many ways to construct the propagator MPO that
often involve a trade off between the maximum time-step and bond dimension. The
development of optimal propagator MPOs has been an object of intense research
over the years~\cite{paeckelTimeevolutionMethodsMatrixproduct2019,
Zaletel2015a, haegemanTimeDependentVariationalPrinciple2011,
yangTimedependentVariationalPrinciple2020}. In this work, we use a modified
second-order Suzuki-Trotter split propagator MPO that is commonly used with the
(second-order) time-evolved block decimation method
(TEBD)~\cite{whiteRealTimeEvolutionUsing2004,
daleyTimedependentDensitymatrixRenormalizationgroup2004,
vidalEfficientSimulationOneDimensional2004}. These MPOs are generally used
to simulate systems with nearest neighbor couplings. For many photosynthetic
systems, the ring topology is biologically relevant. As discussed earlier, the
ring Hamiltonian has an extra coupling, $H_{1P}$ between the
$1$\textsuperscript{st} and the $P$\textsuperscript{th} sites. Now, the
propagator element between two points $S^+_n$ and $S^+_{n+1}$ for the ring can
be written as
\begin{align}
    U_{\text{ring}} & (S^+_n,S^+_{n+1},\Delta t) \approx\nonumber
    \\ & \mel{S^+_{n+1}}{e^{-i H_{1P} \Delta t/{2\hbar}} U_{\text{chain}}
    (\Delta t) e^{-i H_{1P} \Delta t/{2\hbar}}}{S^+_{n}},
\end{align}
where $U_{\text{chain}}$ is the standard second-order TEBD propagator for the
chain. The resulting propagator MPO for the ring is obtained by multiplying the
MPOs corresponding to the ``long bond,'' $e^{-iH_{1P}\Delta t/2\hbar}$, together
with that of the chain. It is feasible to construct the propagator in a
cylindrical form, reflecting the true ring symmetry of the system. However,
there may be other performance concerns. This would be evaluated in a future
work.

%\textcolor{red}{While in the current work a simple
%TEBD-based decomposition is being used, that is not the only formulation
%possible. For long-ranged interactions, the recently introduced
%W\textsuperscript{I,II} can generate very compact representations of the
%propagator~\cite{Zaletel2015a}. Additionally, the time-dependent variational
%principle~\cite{haegemanTimeDependentVariationalPrinciple2011,
%yangTimedependentVariationalPrinciple2020} approach has been shown to be useful
%in propagating wave functions under Hamiltonians with long-ranged interactions.
%Such an approach can be used to also generate the forward-backward propagator.}

The approach discussed till now is a density matrix version of methods like
TEBD or time-dependent DMRG. The next step is to incorporate the effects of the
solvent by accounting for the time non-locality of the influence functional.
Since the different time points can no longer be uncoupled, traditional time
step iteration is impossible. In principle, this would cause the computational
complexity to increase exponentially with the number of time steps; however,
this cost can be avoided by performing an additional DMRG-like tensor
decomposition along the time axis~\cite{boseTensorNetworkRepresentation2021,
boseMultisiteDecompositionTensor2022}. The resulting 2D tensor network, factored
in both space and time, forms the foundation of MS-TNPI.

To construct the MS-TNPI network, one starts with the MPO representations of the
forward-backward propagator between each of the time-points and uses them to
construct a fully factorized tensor network description of the bare path
amplitude tensor $P^{(0)}$:
\begin{align}
    P^{(0)}_{S_{0}^{\pm}\cdots S_{N}^{\pm}} & = \sum_{\left\{\beta_{n}\right\}} \mathbb{T}^{S_{0}^{\pm}}_{\beta_{0}}
    \cdots \mathbb{T}^{S_{n}^{\pm}}_{\beta_{n-1}, \beta_{n}}
    \cdots\mathbb{T}_{\beta_{N-1}}^{S_{N}^{\pm}}.\label{eq:MS-PA_MPS}
\end{align}
Here, each $\mathbb{T}$ is in a matrix product representation decomposed along
the site axis, and $\beta_n$ is the bond dimension connecting the tensors at
time-point $n$ to the one at $n+1$. These indices are called ``temporal'' bonds
because they connect points on the same site but different times. The 2D
structure of the MS-TNPI network is demonstrated in Fig.~\ref{fig:ms-tnpi}.
(For convenience, a more detailed derivation of the tensors that make up
$\mathbb{T}$ is provided in Appendix~\ref{app:MS-TNPI-Network}.) Though
Eq.~\ref{eq:MS-PA_MPS} has been written for the bare path amplitude tensor, the
full tensor also would have a practically identical form. The main difference
being whether the influence functional has been incorporated.

\begin{figure}
    \includegraphics[scale=0.2]{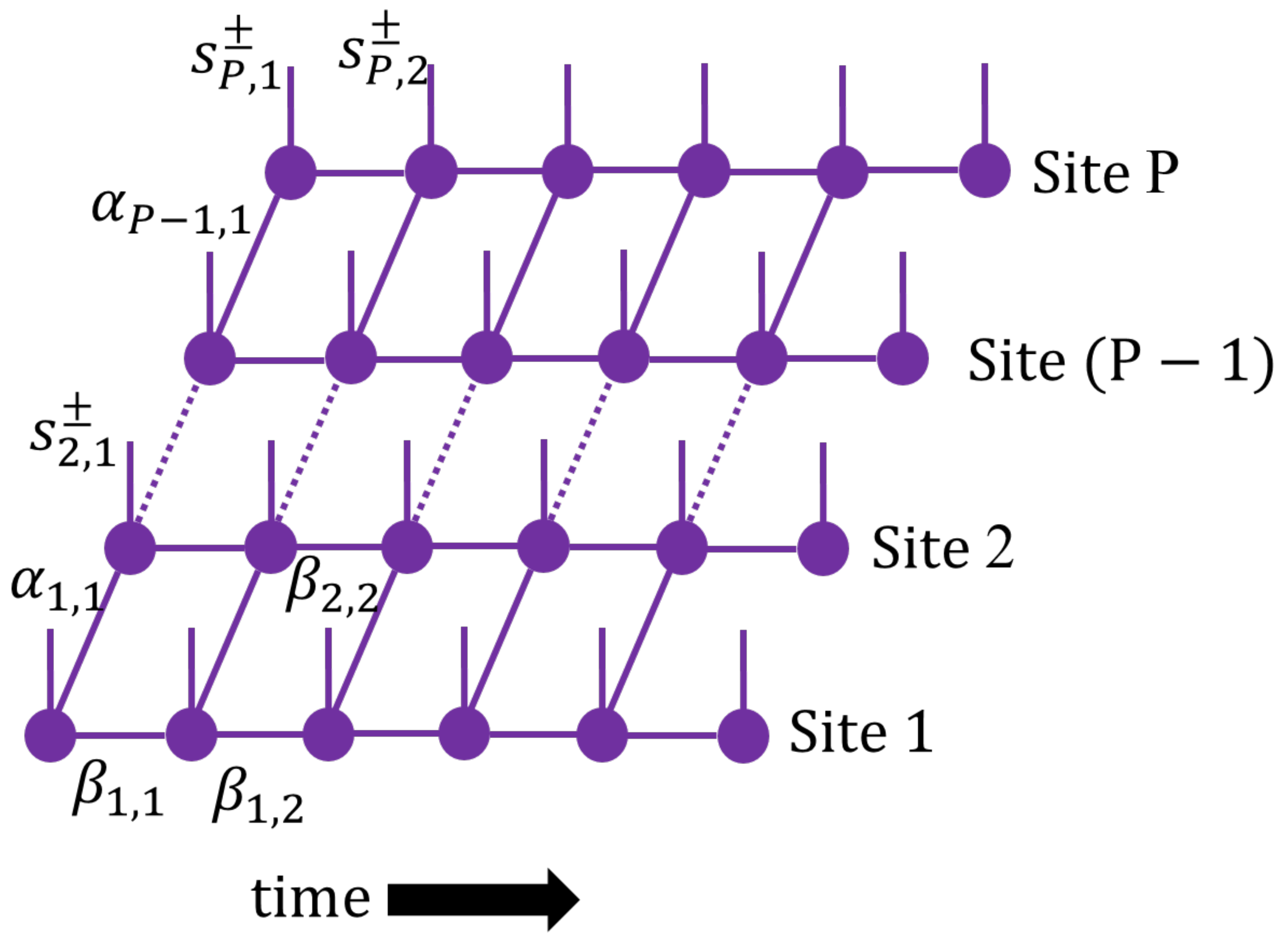}
    \caption{2D MS-TNPI tensor network.}\label{fig:ms-tnpi}
\end{figure}

Since the influence functional has not yet been applied, contracting the 2D
MS-TNPI network, as it stands right now, yields the time-evolved density matrix
for the isolated extended system expressed in the form of an MPS. Using the
tensor network path integral~\cite{boseTensorNetworkRepresentation2021}, one can
define the influence functional MPO for each site or monomer unit and apply it
to the corresponding site as illustrated in Fig.~\ref{fig:MS-TNPI-influence}.
The relevant equations for the influence functional MPO have been summarized in
Appendix~\ref{app:IF-MPO}. Now, upon contraction, the network gives the
resulting time-evolved reduced density matrix. (Though, we have presumed a lack
of correlation between the baths on different sites, it is possible to extend
the structure to take correlation effects into account as well by applying
operators that connect the ``rows'' corresponding to the different sites. In
absence of those correlation effects, the influence functional MPOs directly
affect only the temporal bond dimension and not the spatial bond dimension.)

\begin{figure}
    \includegraphics[scale=0.2]{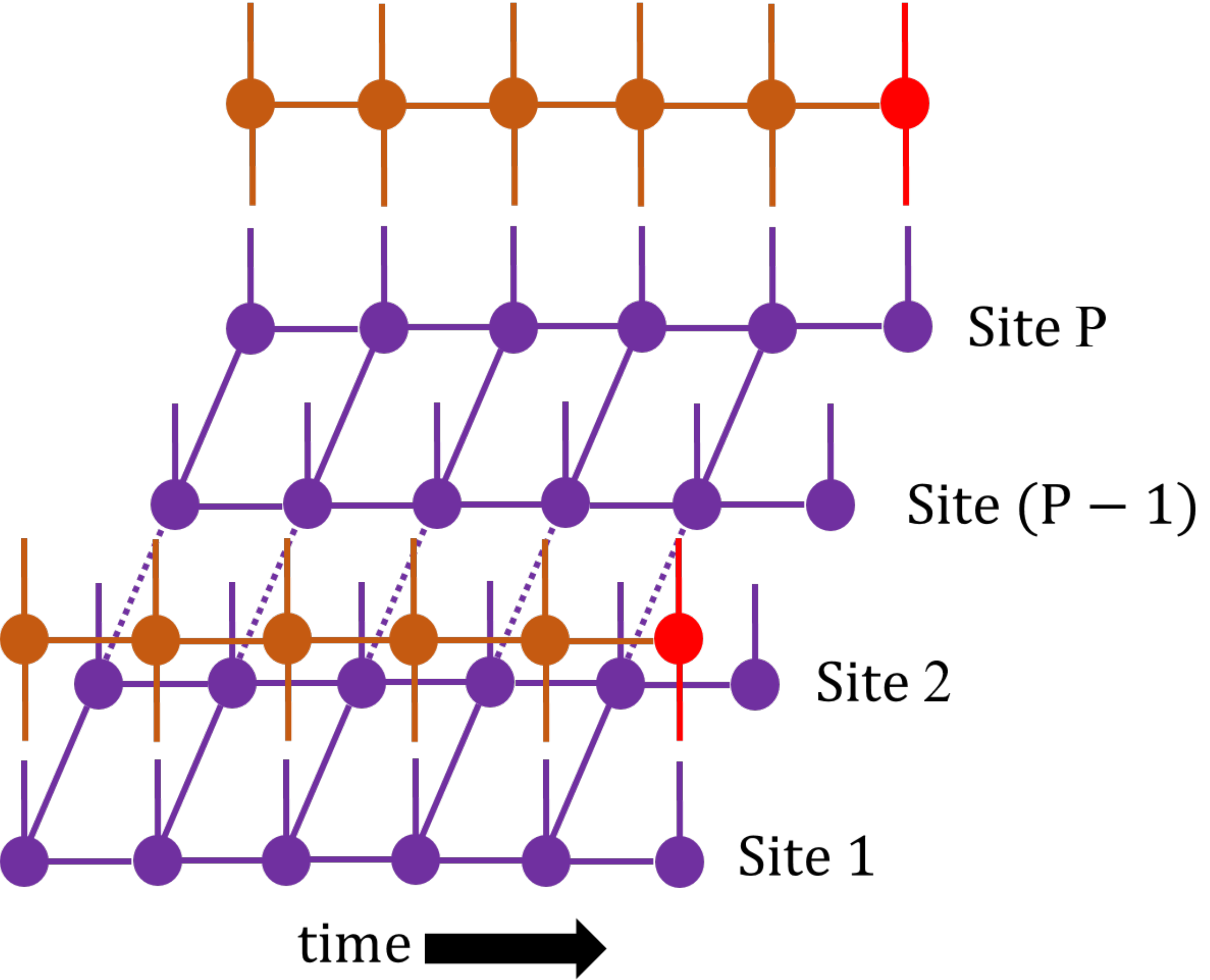}
    \caption{Schematic depiction of the application of the influence functional
    MPOs to selected sites of the extended
system.}\label{fig:MS-TNPI-influence}
\end{figure}

Let us examine the 2D MS-TNPI network corresponding to the path amplitude
tensor in more depth. Along the time axis, it consists of the local path
amplitude tensor for each of the sites, effectively generalizing the TNPI
structure to multiple sites. On the other hand, along the space axis, each of the ``columns''
represents the full state of the system at that time point and is effectively a
generalization of the reduced density matrix as propagated by time-dependent
DMRG methods. The network allows the possibility of many different algorithms
for contracting it. While the present work uses a contraction scheme that
preserves the columns, and consequently obtains the entire time propagated
density matrix as an MPS, future explorations could yield other interesting and
performant schemes.

%The ``bonds'' along the
%time axis joins the system states at different time points and are called
%the ``temporal bonds.'' Similarly, the bonds connecting different sites or
%units at the same time point are called ``spatial bonds.''

Na\"ively speaking, the network should have one column for each time point of propagation. This is due to the
non-Markovian memory induced by the bath. This represents the growth of
computational cost with the propagation time. Much of this exponential
complexity is already controlled through the tensor network decomposition and
the accompanying truncated singular value decomposition filtration schemes.
However, this is not enough by itself. It is well-known that in condensed
phase environments, the memory dies away with the temporal distance between two
points. It is, therefore possible to truncate the memory length to say $L$
time-steps and use $L$ as a convergence parameter. This is achieved through a
procedure for iterative propagation of the density matrix for the extended
system.

\begin{figure}
    \includegraphics[scale=0.2]{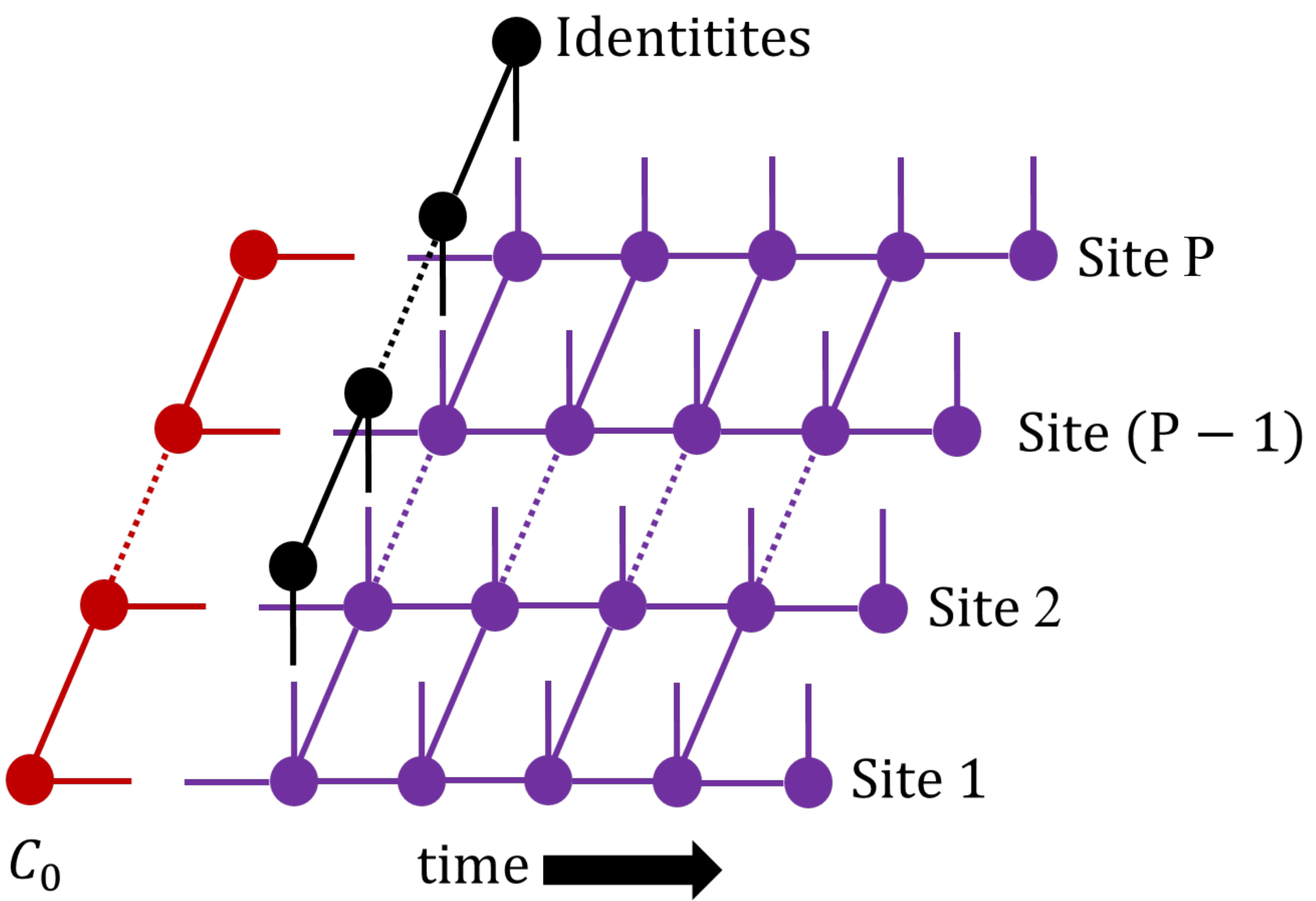}
    \caption{Schematic illustrating the iteration scheme in the MS-TNPI framework.}\label{fig:iter}
\end{figure}

When iteration starts, there are $L$ time-steps and, consequently, $L+1$ columns in
the MS-TNPI network. These are labeled as $C_n$ for $1\le n\le L+1$. For the
initial step of iteration, let $C_0 = \tilde\rho(0)$ written as an MPS. (If, as
in Sec.~\ref{sec:spectra},  we are simulating a correlation function $C(t) =
\Tr[U(t)\,\rho(0)\,A\,U^\dag(t)\,B]$, $C_0 = \tilde\rho(0)\,A$.) The iteration
method can be summarized in the following series of steps. (An MPO-MPS
multiplication is written as $\otimes$ in the following steps.)
\begin{enumerate}
    \item Update $C_0$ by multiplying it by the MPO $C_1$. $C_0 \leftarrow C_1
     \otimes C_0$. (Note that the first column, $C_1$, is an MPO and the
     resulting $C_0$ an MPS.)
    \item Slide all the columns back by one step. $C_j \leftarrow C_{j+1}$ for
     $j<L$. (Now, the new first column is no longer an MPO and has two temporal
     bond indices.)
    \item Update $C_L$ and insert the $C_{L+1}$ to account for the propagator
        between the penultimate and the last time steps. (The working equations
        are in Appendix~\ref{app:MS-TNPI-Network}.) \item Apply the influence
        functional MPO to each row.
    \item Trace over the site indices of $C_1$ to turn it into an MPO again.
\end{enumerate}
Steps (1) -- (5) are repeated as many times as required to get the full
dynamics. This procedure is schematically represented in Fig.~\ref{fig:iter}.

These calculations involve manipulation of high-dimensioned tensors factorized
in different forms. The cost of applying the IF MPO is
$\mathcal{O}\left(m_{t}^3w_{p}^2w_{I}^2d^2\right)$, and the cost of the
contraction is $\mathcal{O}\left(m^3w_{p}^2m_{t}\right)$. Here, $m_{t}$ is the
maximum temporal bond dimension, $m$ is the maximum bond dimension of the
contracting MPS, $w_{I}$ is the maximum bond dimension of the IF MPO. The
maximum bond dimension of the forward-backward propagator of the bare system is
denoted by $w_{p}$ and $d=2$ is the dimensionality of a typical system site.
The computational cost is linear in the number of sites or system size, which
appears as a prefactor in the formal scaling expressions. Though the magnitude
of $m_t$ might be dependent on the memory length, $L$, the exponential growth
of complexity within memory is effectively curtailed. The cost propagation
beyond the memory span of $\tau = L\Delta t$ is strictly linearly proportional
to the number of steps of dynamics beyond the memory length simulated. The
dominant cost is determined by the particular parameters under consideration.
For more details about the contractions involved, please consult
Ref.~\cite{boseMultisiteDecompositionTensor2022}.

\subsection{Absorption Spectrum}\label{sec:spectra}

Absorption spectra are calculated as the Fourier transform of dipole-dipole time
correlation function,
\begin{align}
    \sigma(\omega) & \propto \Re \int_0^t e^{i\omega t} C(t)\dd{t}                       \\
    C(t)           & = \Tr\left(\hat\mu(t)\hat\mu(0)\rho(0)\right).\label{eq:mu_mu_corr}
\end{align}
Here, $\hat\mu(t) = \sum_{i=1}^P \hat U^\dag(t)\hat\mu \hat U(t)$ is the
time-evolved total dipole operator, $\hat\mu = \sum_{i=1}^P \hat\mu_i$. Generally
speaking, the local dipole operators do not point in the same direction.
Therefore, using the site-local basis, $\hat\mu_i = \vec{d}_i
(\dyad{\phi^g_i}{\phi^e_i} + \dyad{\phi^e_i}{\phi^g_i})$, where $\vec{d}_i$ is
the dipole moment vector corresponding to the $i$\textsuperscript{th} unit. For
calculating the absorption spectrum,
\begin{align}
    \rho(0)       & = \tilde\rho(0)\otimes\prod_i \frac{\exp(-\beta V_i)}{Z_i} \\
    \tilde\rho(0) & = \prod^\otimes_i \dyad{\phi^g_i},
\end{align}
where the vibrational manifold associated with the $i$\textsuperscript{th} unit
is distributed thermally at an inverse temperature, $\beta=\tfrac{1}{k_B T}$, on
the ground Born-Oppenheimer surface. The partition function for this
distribution is given by $Z_i$.

In the path integral notation of Sec.~\ref{sec:PI_Exp}, the correlation
function, Eq.~\ref{eq:mu_mu_corr}, can be written as
\begin{align}
&C(N \Delta t) = \sum_{S_{0}^{\pm}\ldots S_{N}^{\pm}}\,\sum_{S_{0}^{'\pm}}\,\sum_{S_{N}^{'\pm}}\,\nonumber\\ 
&\delta_{S_{N}^{'+}, S_{N}^{'-}}\,\tilde{\rho}\left(S_0^\pm,0\right)\,\hat{\mu}\left(S_0^{\pm},S_0^{'\pm}\right)\,P_{S_{0}^{'\pm}\cdots S_{N}^{\pm}}\,\hat{\mu}\left(S_N^{\pm},S_N^{'\pm}\right),\label{eq:mu_mu_corr_PI}
\end{align}
where $\hat{\mu}\left(S_n^{\pm},S_n^{'\pm}\right) =
\mel{S^{'+}_n}{\hat{\mu}}{S^{+}_n} \mel{S^{-}_n}{\mathbb{I}_n}{S^{'-}_n}$ and
$\mathbb{I}_n$ represents the identity operator of the full forward-backward
space at $n$\textsuperscript{th} time point. It is worth noting that in this
case, the total dipole operator only acts on the forward space; furthermore, it
is possible to analytically represent it as an MPO, which is given in
Appendix~\ref{app:Dipole_MPO}. Interestingly, in this form, the MPO is extremely
compact, having a bond dimension of just two. Since we have an MPO expression
for the total dipole operator, the correlation function can be efficiently
computed by applying this MPO to the initial density MPS as well as the final
time-propagated MPS before taking the trace, as per Eq.~\ref{eq:mu_mu_corr_PI}.
By applying the total dipole moment MPO at multiple intermediate points, it
should also be possible to calculate higher order response functions at minimal
extra computational cost.

From here, we could proceed directly to computing the correlation function.
However, the high frequency nature of the electronic absorption necessitates the
use of very short time steps, thereby increasing the memory length.
Additionally, these correlation functions decay slowly which means the total
number of simulation steps would be fairly large. Together, these two factors
serve to increase the computational complexity of these simulations. Such a
direct approach is, therefore, rather inefficient. Fortunately, though, in
this case, it is possible to transform the entire problem into a numerically
simpler problem.

This transformation begins by shifting each site by a constant energy term,
$\bar{E}$, that has the same order of magnitude as the electronic excitation
energy ($\bar{E}\approx \SI{12000}{\per\cm}$ for B850). The system Hamiltonian,
Eq.~\ref{eq:frenkel}, can then be rewritten as
$\hat{H}_{0}=\hat{\bar{H}}_{0}+\hat{\mathcal{H}}_{0}$, where
\begin{align}
    &\hat{\bar{H}}_{0}   = \bar{E}\sum_{i=1}^{P}\dyad{e_{i}}\quad\text{and}\\
    &\hat{\mathcal{H}}_{0}  = \sum_{i=1}^{P}\epsilon_{i}\dyad{e_{i}} + \sum_{i=1}^{P-1} J_i \left(\dyad{e_{i+1}}{e_i} + \dyad{e_i}{e_{i+1}} \right),
\end{align}
with $\epsilon_{i} = E_{i} - \bar{E}$. $\hat{\bar{H}}_{0}$ commutes with
$\hat{\mathcal{H}}_{0}$, so we can factor
$\hat{\bar{U}}=\exp(-\tfrac{i}{\hbar}\hat{\bar{H}}_{0} t)$ out of the propagator
without incurring any Trotter error. Next, we identify the ``bra'' side of
$\hat\mu(0)\tilde\rho(0)$ (corresponding to the backward path) with the
electronic ground state. On the ``ket'' side, which corresponds to the forward
path, it is in a state with a single excitation. Due to the block-diagonal
structure of the Frenkel Hamiltonian, the bra of the time-evolving operator
remains in the ground state and the ket remains in the manifold of singly
excited states. In other words, the backward path remains in the ground state,
and forward path only populates the first excited subspace. In this subspace,
$\hat{\bar{H}}_{0}=\bar{E}$ and $\hat{\bar{U}}=e^{-i \bar{E}t/\hbar}$ Thus,
we have:
\begin{align}
    C(t) & = e^{-i \bar{E} t/\hbar} \mathcal{C}(t),\label{eq:factor_corr}
\end{align}
where $\mathcal{C}(t)$ is the dipole moment autocorrelation function obtained
using the propagator $\hat{\mathcal{U}}$ corresponding to
$\hat{\mathcal{H}}_{0}$. Because the extremely high frequency oscillations have
been factorized out, the time-steps can now be larger. The multiplication by
the fast rotating phase is done as a post-processing step and is equivalent to
a shift of the absorption lineshape to account for the redefining of the zero
of energy. It is worth noting that since the backward path remains in the
ground state, we could have derived a more compressed representation of the
influence functional MPO. While this optimization was not needed here, it may
be required in the future.  

Before concluding this subsection, it is instructive to explore the form
of the ``initial state''. We note that though $\tilde\rho(0)$ is a separable
state, the initial state, defined as the product with the direct sum of the site
local $\hat\mu_j$ operators, is surely not separable.
\begin{align}
    \hat\mu(0)\tilde\rho(0) & = \sum_j \hat\mu_j\tilde\rho(0)                  
                         \\ & = \sum_j \vec{d}_j \prod_{k\ne j}
              I_k\otimes \left(\dyad{\phi^g_j}{\phi^e_j}+\dyad{\phi^e_j}{\phi^g_j}\right) \prod_i \dyad{\phi^g_i} \\
                            & = \sum_j\vec{d}_j \prod_{k\ne j} \dyad{\phi^g_k} \otimes \dyad{\phi^e_j}{\phi^g_j}\label{eq:sumMPS}
\end{align}
where $I_k$ is the identity operator on the $k$\textsuperscript{th} site. Each
of the operators in the summand of Eq.~\ref{eq:sumMPS} is in a direct product
form. This sum over multiple such operators causes the sites to be entangled and
the effective initial state to be non-separable. While we have written out the
equation explicitly for the absorption spectrum, this issue of non-separability
and entanglement of the initial condition is a consequence of the operators
involved in the correlation function. This feature is common to most spectra of
interest. In fact, for the emission spectrum, the initial condition has even
greater entanglement. To our knowledge, MPI is the only other method that is
able to use influence functionals for general extended quantum systems; however,
since it treats the system sites sequentially, it is not designed to handle
non-separable initial states.  This means that the simulation would require
separate runs, each corresponding to a different term in the sum. The fact that
the final result comes from a trace over a non-direct product operator further
increases the number of runs that would be be required. Thus, an MPI calculation
of the correlation function, given by Eq.~\ref{eq:mu_mu_corr}, is likely many
times more costly than a simple calculation of the population dynamics. On the
other hand, since MS-TNPI is compatible with the MPS/MPO framework, and the
total dipole operator can be expressed as an extremely compact MPO; it can
calculate the correlation function at practically the same cost as the
population dynamics.

\subsection{A Note About Convergence}\label{sec:converge}

As with any numerical approach, the simulations included in this work involve a
variety of different convergence parameters. Here, we give a brief description
of the key parameters as well as a quick outline of the procedure used. Loosely
speaking, these parameters can be grouped into two categories: those arising
from the path integral (i.e., time-step and memory length) and those coming
from the SVD compression of the network.  Under our compression scheme, the
singular values, $\lambda_{n}$, are discarded such that
\begin{align}
  \frac{\sum_{n\in\text{discarded}}\lambda_{n}^{2}}{\sum_{n}\lambda_{n}^{2}}<\chi.
\end{align}
The particular value of truncation threshold, $\chi$, used depends on the part
of the network being compressed.  We used two different truncation thresholds,
here, $\chi_t$ (used for compressing the bonds along the temporal axis) and
$\chi_s$ (for the spatial axis). Conceptually, the value of $\chi_t$ changes
with the memory length, $L$, while the value of $\chi_s$ changes with the number
of sites and the strength of the couplings between them. Of course, while doing
the calculation, these clean conceptual divisions do not hold and the two
dimensions start affecting each other.

Typically, one starts by choosing a particular value of time-step and $L$, and
iterating through different cutoffs to achieve convergence with respect to
them. Subsequently, the time-step and $L$ are changed, repeating the process of
converging the cutoffs at each step, to find the largest converged time-step
and the smallest $L$. Unlike typical system-solvent decomposed methods, here,
the Trotter error caused by the time-step stems from both the system-solvent
split as well as the system-system split. The memory length, $L$, however is
only caused by the local baths. 

\section{Results}\label{sec:results}

\begin{figure}
    \includegraphics[scale=0.3]{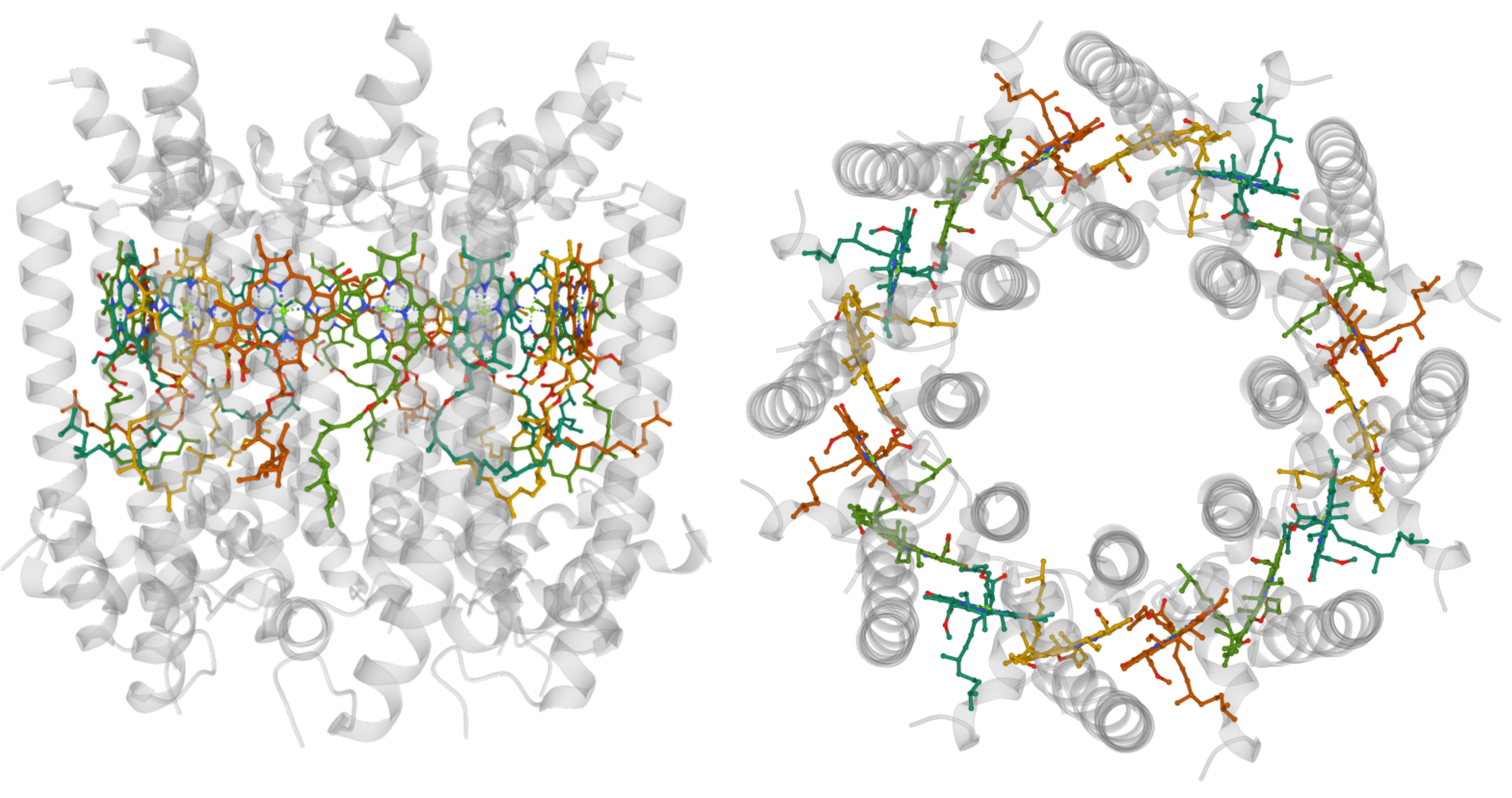}
    \caption{Side and top views of the B850 ring of LH2. The molecular vibrations and the protein scaffolding affect the dynamics.}\label{fig:b850}
\end{figure}

\begin{figure}
     \includegraphics{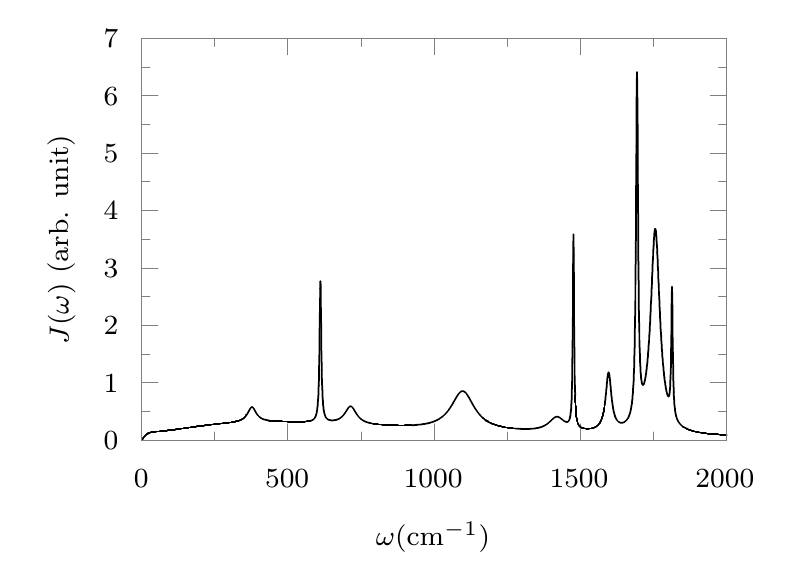}
     \caption{Spectral density corresponding to the B850 ring as calculated in
 Ref.~\cite{olbrichTimeDependentAtomisticView2010}.}\label{fig:spect_dens}
\end{figure}

The B850 ring of LH2, shown in Fig.~\ref{fig:b850}, is an important component of
photosynthetic complexes. It has been previously studied with approximate
spectral densities. Here, we use the accurate spectral densities derived
by~\citet{olbrichTimeDependentAtomisticView2010} to model the interaction of the
system with the rigid molecular vibrations and the impact of the protein
scaffolding. The resultant spectral density obtained along MD trajectories with
ZINDO/S-CIS calculations for the energy gap is shown in
Fig.~\ref{fig:spect_dens}. It is well-known that the B850 ring can be decomposed
into constituent dimers with high intra-dimer electronic couplings. The
couplings between the different dimers is considerably smaller. The electronic
couplings between the nearest neighbors, calculated using the method of
transition charges from electrostatic potentials
(TrEsp),~\cite{rengerTheoryExcitationEnergyTransfer2009,
madjetIntermolecularCoulombCouplings2006} alternate between
\SIlist{173;140}{\per\centi\meter}~\cite{olbrichTimeDependentAtomisticView2010}.
Notably, these values are significantly less than ones derived from experiments
because they take environmental screening effects into account. As a point of
comparison and to understand the system better, we also consider the
experimentally derived electronic couplings of \SIlist{315;245}{\per\cm} as
reported by~\citet{freibergExcitedStateDynamics2009} in their experimentally fit
models. These parameters have been used
by~\citet{strumpferEffectCorrelatedBath2011} in their study of the dynamics of
B850 ring coupled with a Drude-Lorentz spectral density. Other experimentally
derived numbers~\cite{tretiakBacteriochlorophyllCarotenoid2000} are also of
similar magnitude.

\begin{figure}
    \subfloat[Excited state populations of each of the BChl sites.]{\includegraphics{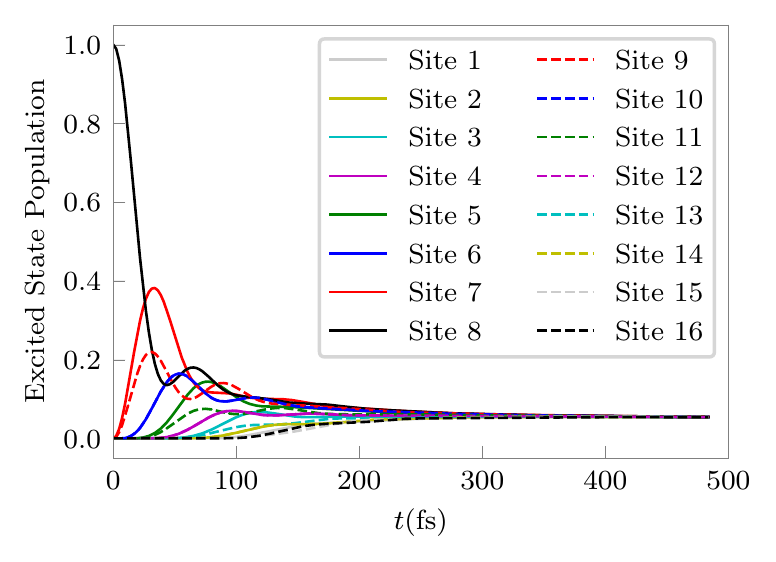}}

    \subfloat[2D plot of the transfer of excitations from one BChl site to
    another.]{\includegraphics{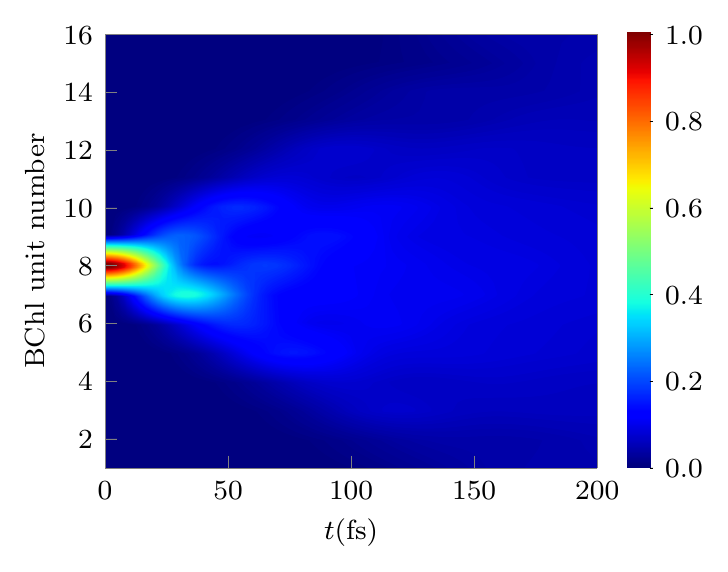}}

    \caption{Excited states of the bacteriochlorophyll corresponding to the
    B850 ring with the TrEsp couplings.}\label{fig:popln_kleinekathoefer}
\end{figure}

\begin{figure}
    \subfloat[Excited state populations of each of the BChl sites.]{\includegraphics{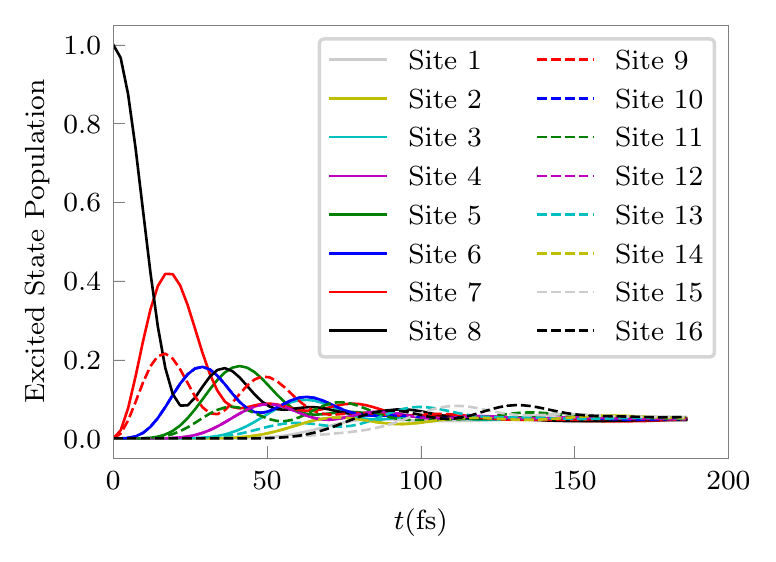}}

    \subfloat[2D plot of the transfer of excitations from one site to
    another.]{\includegraphics{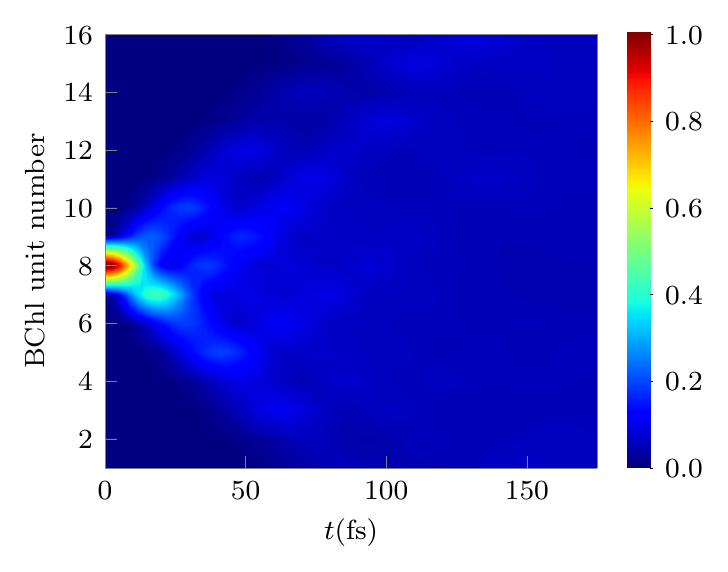}}

    \caption{Excited states of the bacteriochlorophyll corresponding to the B850
    ring with the experimentally derived couplings.}\label{fig:popln_experiment}
\end{figure}

The dynamics of exciton transport in the B850 ring corresponding to the initial
excitation of the 8\textsuperscript{th} BChl unit ($\tilde\rho(0) =
\dyad{e_8}$) with the TrEsp couplings is shown in
Fig.~\ref{fig:popln_kleinekathoefer}. For the calculations shown in here,
typically $\Delta t = \SI{4.84}{\fs}$ yields converged results with a memory
span of $L\Delta t = \SI{24.19}{\fs}$. The cutoffs for this problem are very
different along the temporal and spatial axes. Along the time-axis, $\chi_t$
converged around $10^{-10}$ whereas the spatial cutoff, $\chi_s$ was converged
around $10^{-5}$ -- $10^{-7}$. A representative converged simulation of the
full dynamics takes around 6 hours on an Intel\textregistered{}
Xeon\textregistered{} Gold CPU. The runtime is, of course, extremely dependent
on the exact parameters of the system under study and the levels of singular
value decomposition truncation that is being done. Because of the low
couplings, the first peak of the initially excited site happens at $\approx
\SI{75}{\femto\second}$, which is significantly later than what is expected
from experimentally derived couplings. As a comparison, we demonstrate the
corresponding dynamics of B850 ring parameterized by experimentally derived
electronic couplings in Fig.~\ref{fig:popln_experiment}. Notice that in this
case, the prominent hump in the excited state population of the initially
excited monomer happens around $\SI{37}{\femto\second}$. 


The dynamics of the dimerized ring of identical chlorophyll molecules displays
an interesting symmetry. Consider all pairs of BChl units equidistant from the
initially excited one, which in this case is the 8\textsuperscript{th} unit.
If the electronic couplings between all nearest neighbor pairs were equal, the
dynamics of the monomers of any pair would have been identical. However,
because of the alternating nature of these couplings, such a symmetry would be
absent. Interestingly, this situation leads to a different symmetry. Now it is
every alternate pair that has identical dynamics and the other pairs have
different dynamics. Of course, because the number of units in this case is
even, the unit diametrically opposite to the initially excited unit, the
8\textsuperscript{th} BChl unit in this case, is unique.

\begin{figure}
    \subfloat[BChl ring with average intermonomer electronic couplings.
    Excited state population of site 7 is same as that of site 9, site 6 is
    equal to site 10, so on.]{\includegraphics{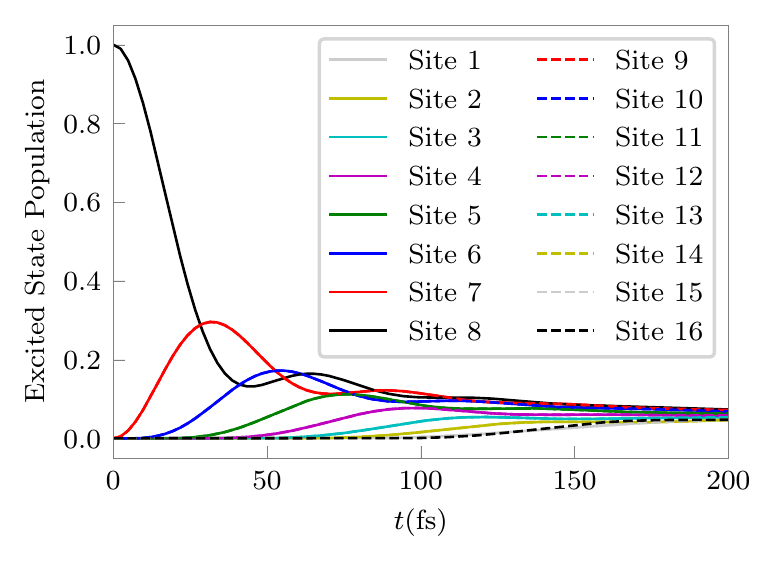}}

    \subfloat[Dimerized BChl ring with electronic couplings obtained using
    TrEsp. Data is the same as Fig.~\ref{fig:popln_kleinekathoefer} but shown
    for a shorter time duration.]{\includegraphics{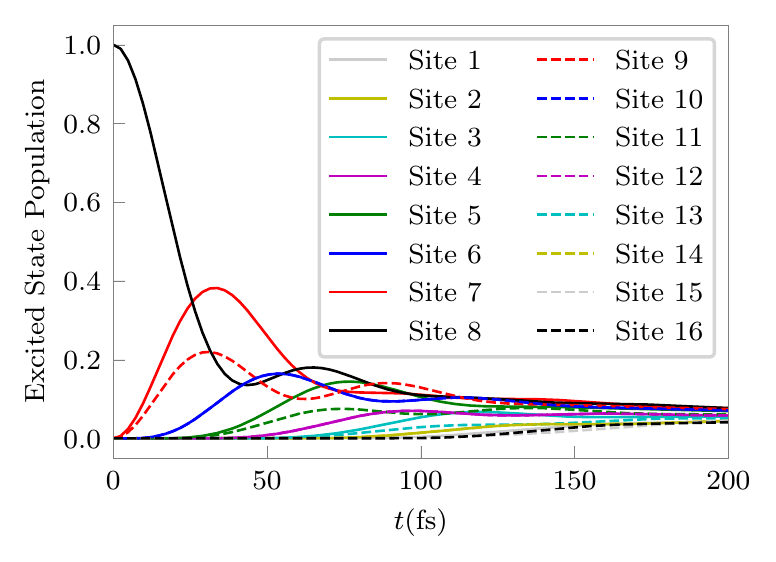}}

    \caption{Dynamics for BChl rings with identical and alternating electronic
        couplings.}\label{fig:popln_comp}
\end{figure}

Now, let us compare the dynamics of the dimerized
BChl ring with the TrEsp couplings with a BChl ring with all the couplings set
at the average value of the TrEsp couplings. Figure~\ref{fig:popln_comp} shows
the dynamics corresponding to the non-dimerized (with average couplings,
Fig.~\ref{fig:popln_comp}~(a)) and dimerized (alternating couplings,
Fig.~\ref{fig:popln_comp}~(b)) ring. Note that in
Fig.~\ref{fig:popln_comp}~(a), all the lines but the ones corresponding to 8
and 16 are paired. For example, the dynamics corresponding to the
7\textsuperscript{th} and the 9\textsuperscript{th} sites are identical (they
have the same colors in the figure, but different line styles, so it seems
like there is only one single line). The same applies to the dynamics of the
6\textsuperscript{th} and the 10\textsuperscript{th} sites. However, as
discussed, this is not the case in Fig.~\ref{fig:popln_comp}~(b). For this
case, the dynamics of site 7 and site 9 are different, as is the dynamics of
site 5 and 11. However the dynamics of site 6 is same as that of site 10, as
is the dynamics of sites 4 and 12.

\begin{figure}
    \includegraphics{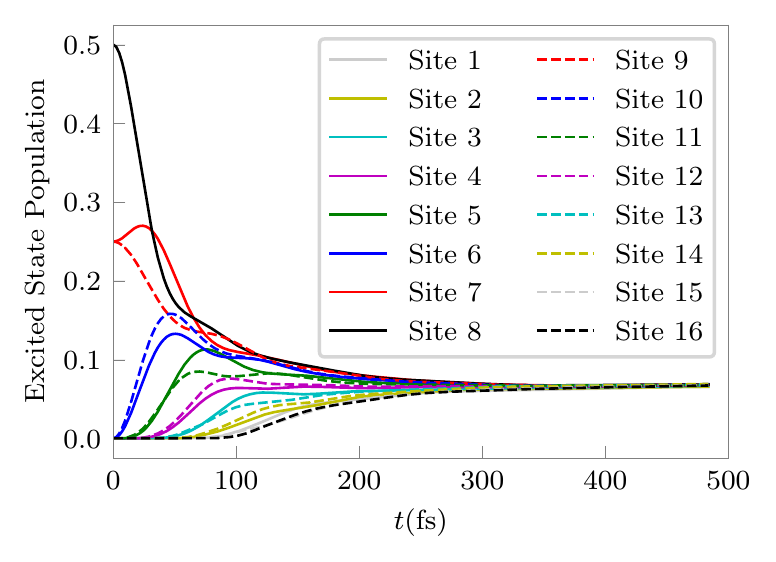}
    \caption{Dynamics starting from $\tilde\rho(0) = 0.5\dyad{e_8} + 0.25\dyad{e_7} + 0.25\dyad{e_9}$ using the TrEsp parameters.}\label{fig:mixed_state}
\end{figure}

To further our understanding, consider the dynamics corresponding to a more
involved initial state. Till now, we have discussed the dynamics following an
excitation of only a single site, the 8\textsuperscript{th} site in our case.
Let us assume that the initial density is defined by $\tilde\rho(0) =
0.5\dyad{e_8} + 0.25\dyad{e_7} + 0.25\dyad{e_9}$. The system is, therefore,
initially in a statistical ensemble with the 7\textsuperscript{th},
8\textsuperscript{th} and 9\textsuperscript{th} sites getting excited with
different probabilities. The dynamics is shown in Fig.~\ref{fig:mixed_state}.
The coupling between the 7\textsuperscript{th} and 8\textsuperscript{th} BChl
units is higher than that between the 8\textsuperscript{th} and
9\textsuperscript{th}. This leads to a transient build-up of excitonic
population in the 7\textsuperscript{th} site at around $\SI{30}{\fs}$, while
the population of the 9\textsuperscript{th} site shows a more or less monotonic
decay. Not only does the 9\textsuperscript{th} BChl unit receive population
from the 8\textsuperscript{th} unit slowly, it also quickly leaks population
into the 10\textsuperscript{th} unit which is completely in the ground state
because of a high electronic coupling.

\begin{figure}
    \includegraphics{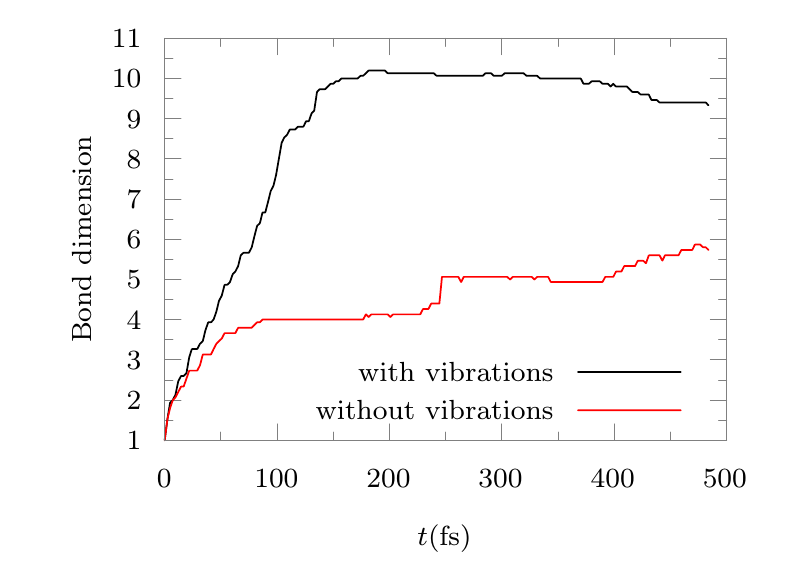}
    \caption{Average bond dimensions of the reduced density MPS for the B850
    ring as a measure of average intersite entanglement in presence and absence
of the vibrational bath.}\label{fig:entanglement}
\end{figure}

A major consideration in multisite systems is the entanglement between the
individual sites. Here, we use the average bond dimension of the reduced
density MPS as a measure of the entanglement. It is intuitively quite clear
that the presence of the bath should change the growth of this bond dimension
and consequently the entanglement between the sites. We have
shown~\cite{boseMultisiteDecompositionTensor2022} that in the case of the Ising
model, the coupling to the local baths severely restricts the growth of the
average bond dimension. In Fig.~\ref{fig:entanglement}, we show the growth of
the average bond dimension for the B850 system both with and without the
presence of the vibrational baths. It is surprising that in this case, the
average bond dimension, and consequently the intersite entanglement, of the
bare B850 system does not really grow and is very small. Additionally, it is
the incorporation of the vibrational bath that leads to an increase in the bond
dimension. Though it must be noted that the bond dimension despite being
greater in presence of the bath, is still quite small. This reversal of
patterns vis-\`a-vis the Ising model is probably unique to the Frenkel model
and might be because of the block diagonal structure of the Hamiltonian. It
might also arise as a consequence of the nature of the quantum transport
process.

\begin{figure}
    \includegraphics{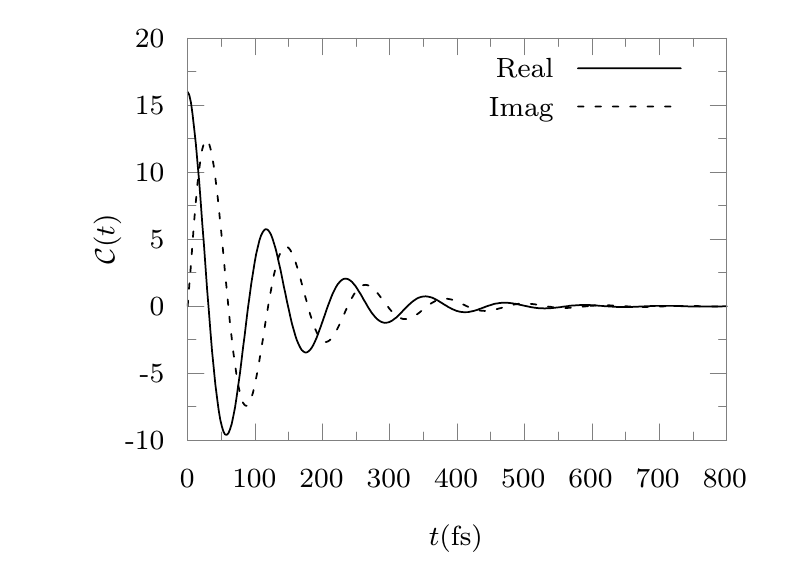}
    \caption{Correlation function after factorizing out the average excitation energy according to Eq.~\ref{eq:factor_corr}.}\label{fig:corr}
\end{figure}

\begin{figure}
    \includegraphics{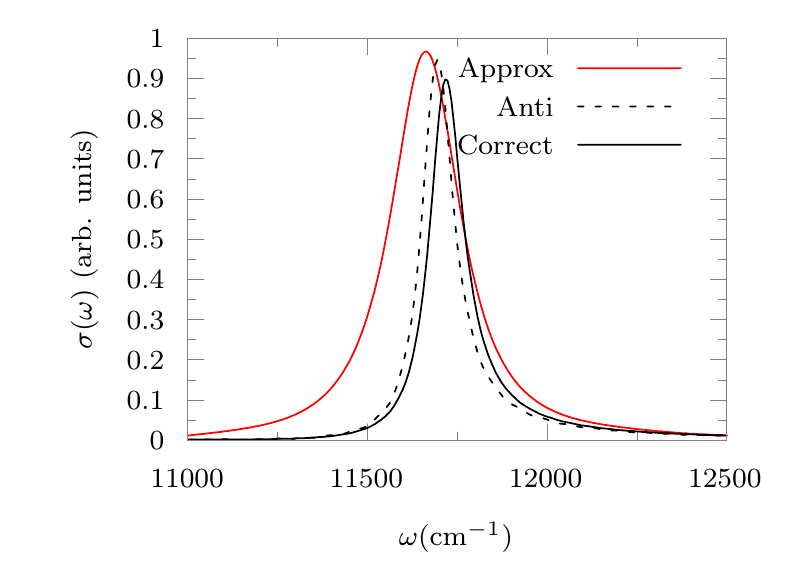}
    \caption{Absorption spectrum corresponding to the TrEsp couplings and constant excitation energies. Black dashed line: antiparallel dipole moments. Black solid line: correct dipole moments. Red solid line: second-order cumulant approximation.}\label{fig:const_exc_spectrum}
\end{figure}

Populations do not give a full account of the dynamics as explored through
experiments. We consider the absorption spectra corresponding to the two
different couplings. As a zeroth order approximation, we first consider the
dipole moment vectors to point in the opposite directions for neighboring
monomers. The convergence parameters for these simulations are mostly the same as
those used for the populations. The most notable difference being the memory
span which is $L\Delta t = \SI{33.86}{\fs}$ here. The correct dipole moment
vectors are tangential to the B850 ring but are oriented in opposite
manners~\cite{Hu1997a, strumpferLightHarvestingComplexII2009}. The mean optical
excitation energy is taken to be
\SI{12098}{\per\cm}~\cite{olbrichTimeDependentAtomisticView2010}. First, we
consider the spectrum corresponding to the TrEsp couplings. The dipole moment
autocorrelation function without the high-frequency oscillations is
demonstrated in Fig.~\ref{fig:corr}. The spectra with the antiparallel dipole
moments and the correct dipole moments are shown in
Fig.~\ref{fig:const_exc_spectrum}. (A representative converged spectrum
calculation takes roughly 2 hours on an Intel\textregistered{}
Xeon\textregistered{} Gold CPU.) Along with the numerically exact MS-TNPI
results, we report an approximate spectrum calculated within the second-order
cumulant approximation~\cite{damjanoviciExcitonsinPhotosynthetic2002}.

\begin{figure}
    \includegraphics[scale=0.5]{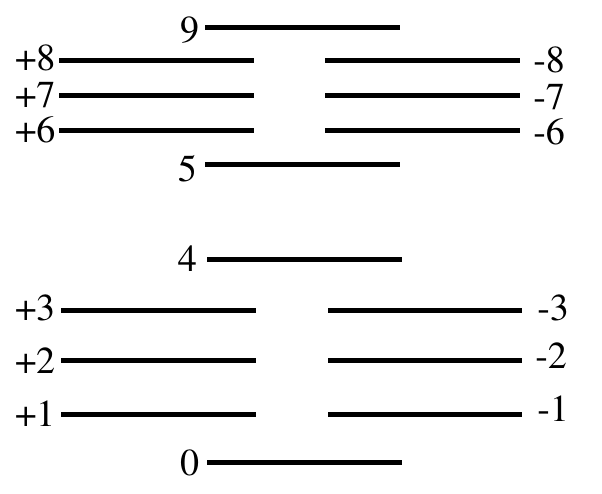}
    \caption{Structure and spacings of the energies of the excitons for a ring with alternating couplings.}\label{fig:exciton}
\end{figure}  

The analysis of the absorption spectrum in terms of the Frenkel excitons is
well-understood. In presence of a system with cylindrical symmetry, there are
two bands of excitons as schematically demonstrated in
Fig.~\ref{fig:exciton}~\cite{Hu1997a}. The exact energies of the excitonic
eigenstates corresponding to the different system parameters is given in
Appendix~\ref{app:exciton}. The degenerate states are labeled as $\ket{\pm k}$.
The gap between the two bands is approximately $2|V_1 - V_2|$, where $V_1$ and
$V_2$ are the two electronic couplings. Therefore, if there is a constant
coupling, the states $\ket{4}$ and $\ket{5}$ would be degenerate as
well~\cite{strumpferLightHarvestingComplexII2009}. For the case where the dipole
moments are antiparallel, the lowest energy exciton, $\ket{0}$, gets excited.
However, it is well-known that the correct dipole moment actually excites into
the degenerate states of $\ket{\pm
1}$~\cite{Hu1997a,strumpferLightHarvestingComplexII2009}. Therefore, as shown in
Fig.~\ref{fig:const_exc_spectrum}, we expect to see a small blue-shift of the
central frequency of the peak corresponding to the correct dipole moments
vis-\`a-vis the antiparallel ones. The excitons are coupled to each other
through interactions with the site-local baths. Thus, the peak is slightly
shifted and significantly broadened in the presence of the dissipative medium.
The second-order cumulant approximation spectrum is quite red-shifted with
respect to the correct dipole moment MS-TNPI spectrum.

\begin{figure}
    \includegraphics{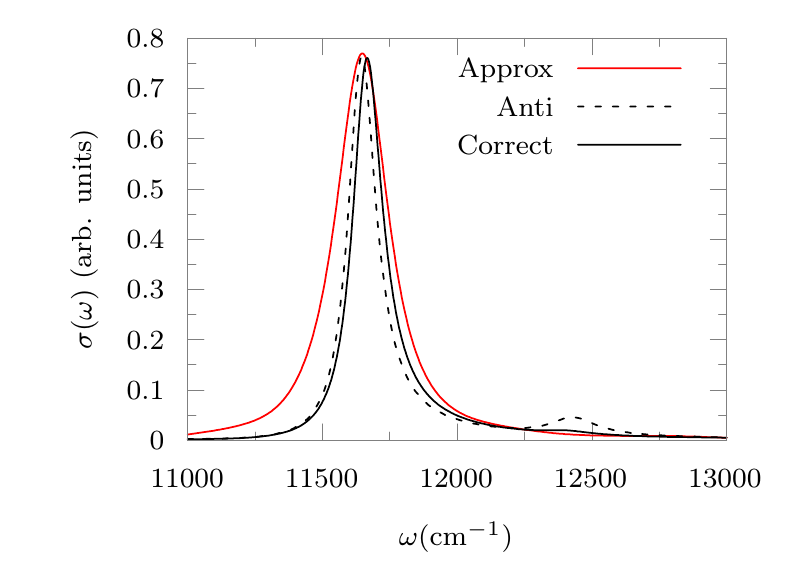}
    \caption{Absorption spectrum corresponding to the TrEsp couplings and varying excitation energies. Black dashed line: antiparallel dipole moments. Black solid line: correct dipole moments. Red solid line: second-order cumulant approximation.}\label{fig:diff_exc_spectrum}
\end{figure}

Now, it is well-known that the electrostatics of the photosynthetic complex
often induces a change in the excitation energies of the monomers. Typically,
the excitation energy alternates with a difference of around
\SI{197}{\per\cm}~\cite{strumpferLightHarvestingComplexII2009}. We have also
simulated and plotted the absorption spectra corresponding to this case in
Fig.~\ref{fig:diff_exc_spectrum}. The incorporation of this asymmetry in the
excitation energy gives rise to a smaller peak close to \SI{12300}{\per\cm}
when using the antiparallel dipole moments. This secondary peak is caused by
excitations into the highest energy exciton, $\ket{9}$, which is now permitted
by the symmetry. For both the correct and antiparallel dipole moments, the main
peak in the spectrum is red-shifted in comparison to the case where the
monomers have the same excitation energy, due to a change in the eigenvalue
spectrum. Also, it is interesting that the agreement with the approximate
spectrum is much better when the varying excitation energies are incorporated.

\begin{figure}
    \includegraphics{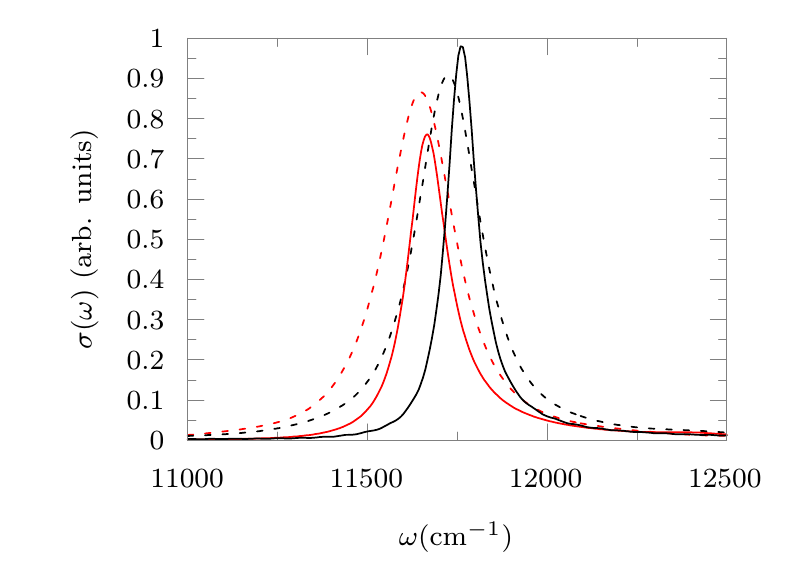}
    \caption{Absorption spectra for TrEsp (red) and experimentally derived parameters (black) with correct dipole moments and asymmetry present. Dashed line: second-order cumulant approximation. Solid line: MS-TNPI spectra.}\label{fig:spectrum_exp}
\end{figure}

As we had mentioned earlier, the electronic couplings obtained via TrEsp are
significantly smaller than the ones derived from experiments. The same holds
for the mean optical excitation energy. We calculate the absorption spectrum
for the B850 ring with a mean optical excitation energy of
$\SI{12390}{\per\cm}$ and couplings of \SIlist{315;245}{\per\cm} as reported
by~\citet{freibergExcitedStateDynamics2009}. The difference between the
excitation energies of consecutive chlorophyll units is once again taken to be
$\SI{197}{\per\cm}$. The comparison of the spectrum corresponding to these
experimental parameters with the one corresponding to the TrEsp parameters is
presented in Fig.~\ref{fig:spectrum_exp}. Clearly, the spectrum corresponding
to the experimentally derived parameters is significantly blue-shifted with
respect to the TrEsp parameters. The peak at $\SI{11763}{\per\cm}$ corresponds
to $\SI{850.12}{\nm}$. This is in comparison to the TrEsp parameter peak at
$\SI{11666}{\per\cm}$ or $\SI{857.19}{\nm}$. Because of the higher electronic
couplings, the damping effect of the bath is less pronounced leading to a
significantly sharper peak. It is interesting that in contrast to the MS-TNPI
spectra, the peak widths of the two approximate spectra are quite similar to
each other. The central frequencies of the approximate peaks have a difference
of roughly $\SI{78}{\per\cm}$ between them, which is smaller than the
difference between the MS-TNPI peaks. The peak of the true experimental
spectrum~\cite{olbrichTimeDependentAtomisticView2010} corresponding B850 region
is red-shifted with respect to the TrEsp peak. This means that the agreement of
the peak corresponding to the higher excitation energy and couplings with the
experiment is worse than the agreement of the TrEsp peaks. Thus the effect of
shielding, that leads to the smaller coupling values, are quite important.

\begin{figure}
    \includegraphics{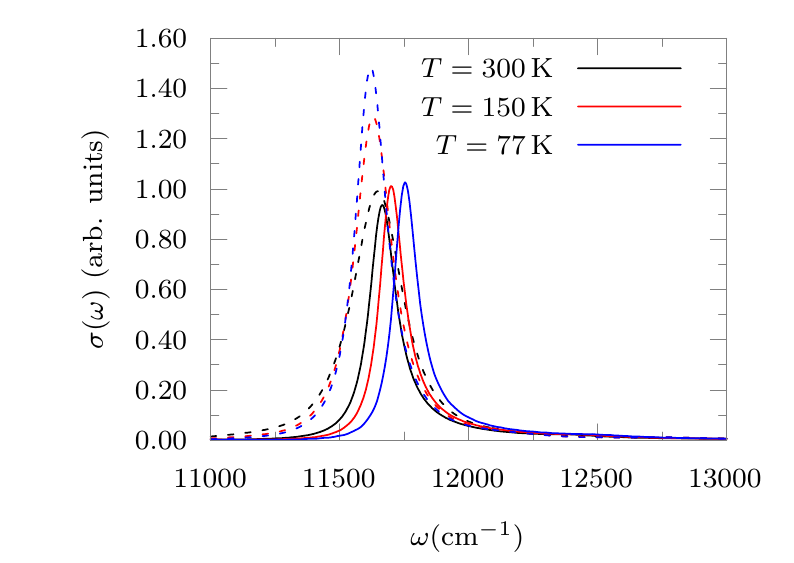}
    \caption{Comparison of spectra at \SIlist{300;150;77}{\kelvin} for the TrEsp parameters with varying excitation energies. Effects of temperature on ``solvent'' spectral density has been ignored for simplicity. Solid lines: MS-TNPI spectra. Dashed lines: second-order cumulant approximation.}\label{fig:spect_temp_comp}
\end{figure}

Finally, we explore how the approximate and exact absorption spectra change
with temperature. It needs to be noted that this exploration has a caveat. The
spectral density coming from the protein and the rigid molecular vibrations is
not necessarily independent of the temperature. The validity of the spectral
density across various temperature regimes would need to be verified on a
case-by-case basis. However, for simplicity, we would keep using the same
spectral density across the temperatures. The spectra are calculated for the
TrEsp parameters with alternating excitation energies at
\SIlist{300;150;77}{\kelvin}. The comparison is shown in
Fig.~\ref{fig:spect_temp_comp}. We see that as the temperature decreases, the
peak of the exact spectrum gets sharper and shows a blue shift. This is
consistent with observations reported by~\citet{chenOpticalLineShapes2009},
though, the magnitude of the blue shift, which is dependent on the exact
parameters and spectral density, is much smaller their case. The approximate
spectrum peaks, on the other hand, shows a red shift while getting sharper as
temperature decreases. This suggests that further investigation into various
approximations for the spectra can give interesting insights into both the
systems under study and the nature of the approximations.

\section{Conclusions}\label{sec:conclusions}

Understanding electronic energy transfer processes is important. However,
studying the dynamics of extended systems with dissipative media is an
extraordinarily challenging problem. System-solvent decomposition is a commonly
used technique to accurately simulate open quantum systems. It handles the
exponential scaling of quantum mechanics by limiting it to a small dimensional
subspace. However, with such extended systems, the exponential scaling of
quantum mechanics is not sufficiently curbed to allow for efficient numerical
simulations. We have recently introduced MS-TNPI to address this problem using
a density matrix renormalization group-like decomposition along with
Feynman-Vernon influence functional. Here, we use it to study EET in a B850
ring of LH2 with vibrational spectral densities obtained using molecular
dynamics. Previous numerically exact studies of the dynamics of such systems
have typically been done with the Drude-Lorentz model spectral density.

In this paper, we have shown how MS-TNPI can be simply extended to account for
the ring structure that is almost ubiquitous in photosynthetic complexes in
purple bacteria. MS-TNPI can efficiently simulate these systems as well. While
we use a ``flat'' 2D structure for simulating the ring system, it is
conceivable that having the 2D structure turned into cylindrical form,
reflecting the actual topology of the system, might bring additional
computational benefits. Such ideas would be explored in the future. We have
also analyzed and massaged the expressions for the absorption spectrum to make
it fit for MS-TNPI. Taking advantage of the availability of the full many-body
reduced density matrix for the extended system, MS-TNPI can efficiently
simulate the required correlation functions and higher order response
functions.

We have shown the impact of the different parameters on the direct EET
dynamics in the B850 ring. The TrEsp couplings with the ZINDO/S-CIS excitation
energies are generally much smaller than typical experimentally derived values.
The dynamics corresponding to both cases have been simulated. The bath has similar
effects on both the parameters. However, owing to the faster oscillations
corresponding to the experimentally derived values, the oscillatory nature
propagates even to the most distant BChl units before getting washed away.
Additionally, subtle effects stemming from the unequal electronic couplings get
amplified when using a more complex initial condition where multiple BChl units
are statistically excited. Future work would focus on studying the impact of
light on B850, taking into consideration effects stemming from the varying
alignments of the site-local dipole vectors and spatial inhomogeneity of the
light-BChl interaction.

Additionally, we have simulated the absorption spectrum, incorporating the full
spectral density, for the ring using various approximations culminating in a
simulation with the most appropriate parameters. The B850 ring is characterized
by non-parallel transition dipole moments and unequal monomer excitation
energies. To better understand the impact of these transition dipole moments,
we started with a very simple zeroth-order approximation where the transition
dipole moments are anti-parallel. We show that consistent with excitonic wave
function-based analysis, if the electronic excitation energies are identical,
there is only a single peak. However, the inhomogeneities induced by the local
electrostatic environment lead to a secondary peak that comes from excitation
into the highest energy excitonic level. Subsequently, we analyse the effect of
using the transition dipole moment with the correct form. Incorporation of the
correct dipole moment operator along with the varying excitation energies still
produces a spectrum that is blue-shifted with respect to experimental
spectra~\cite{olbrichTimeDependentAtomisticView2010}. This is probably due to
inaccurate excitation energies, electronic couplings, and limitations of the
model. These calculations are compared with second order cumulant
approximation~\cite{damjanoviciExcitonsinPhotosynthetic2002}. The approximate
spectra are consistently broader than the MS-TNPI calculations, and generally
slightly red-shifted. It is interesting to note that the effect of the
temperature on the approximate spectra is qualitatively different from that on
the exact spectra. While on decreasing the temperature, the exact spectrum
shifts to higher frequencies, the approximate spectrum shifts to lower
frequencies. Both the types of simulations show the sharpening of the
absorption spectrum at lower temperatures. Consequences of adding static
disorder to the Hamiltonian can be trivially incorporated in the MS-TNPI
procedure through an external Monte Carlo averaging of separate MS-TNPI runs. A
detailed exploration of such effects would be the topic of a future
exploration.

MS-TNPI does not restrict the simulation to the first excitation
subspace as many other methods do. These ``full space'' simulations, however,
still remain quite simple. The singular value decompositions involved seem to
be able to filter out the unnecessary information and lead to very compact
representations. We simulate the dynamics corresponding to a local excitation,
and show that the entanglement between the sites as calculated by the average
bond dimension does not grow exponentially with time, even in the absence of
the bath. This is unlike what happens in say the Ising
model~\cite{boseMultisiteDecompositionTensor2022}, and is probably due to the
sparsity and the block-diagonal structure of the Frenkel Hamiltonian. It is
also interesting that unlike the case of the Ising model where the presence of
the bath controls the entanglement between different
sites~\cite{boseMultisiteDecompositionTensor2022}, in the case of the
Frenkel-Holstein model, the presence of the bath actually serves to slightly
increase the entanglement. This deserves further study.

While here we have explored the dynamics and the absorption spectra, other
experimentally realizable observables can be also be simulated with similar
conceptual simplicity. Further investigation of other observables, especially
multi-time correlation functions and longer ranged interactions will be the
focus of future work. Incorporation of long ranged interactions, through more
advanced propagators (e.g., $W$\textsuperscript{I,II} or TDVP based propagators), would be
important in capturing dipole-dipole interactions between distant monomers in
the Hamiltonian. MS-TNPI provides a flexible scheme
for incorporation of increasingly complex Hamiltonians and effects of baths in
a unified framework, making it a lucrative method for studying quantum
transport in extended quantum systems.

\section*{Acknowledgments}
A.\,B. acknowledges the support of the Computational Chemical Science Center:
Chemistry in Solution and at Interfaces funded by the US Department of Energy
under Award No. DE-SC0019394. P.\,W. acknowledges the Miller Institute for
Basic Research in Science for funding.

\appendix

\section{MS-TNPI Tensor Network}\label{app:MS-TNPI-Network}

Here, we give a short outline of the exact expressions for deriving
the 2D MS-TNPI tensor network~\cite{boseMultisiteDecompositionTensor2022}. The
reduced density matrix of a quantum system coupled to a dissipative bath is
given by the following path integral expression,
\begin{align}
    \tilde\rho(S^\pm_N, N\Delta t) &= \sum_{\{S^\pm_j\}} P^{(0)}_{S^\pm_0,S^\pm_1,\ldots,S^\pm_N} \tilde\rho(S^\pm_0, 0) F\left[\{S^\pm_j\}\right]\\
    P^{(0)}_{S^\pm_0,\ldots,S^\pm_N} &= K(S^\pm_N, S^\pm_{N-1}, \Delta t)\ldots K(S^\pm_1, S^\pm_0, \Delta t)\label{eq:pi}
\end{align}
where $F$ is the Feynman-Vernon influence
functional~\cite{feynmanTheoryGeneralQuantum1963}, $P^{(0)}$ is the bare path
amplitude tensor and $K$ is the forward-backward propagator. The system states
at the $n$\textsuperscript{th} time point are collectively denoted by $S_n^\pm$.
When referring to a specific site and time point, the first index represents the
spatial index and the second one the temporal index (i.e., $s^\pm_{i,n}$
corresponds to the state of the $i$\textsuperscript{th} site at the
$n$\textsuperscript{th} time point). The forward-backward propagator is the
superoperator that evolves the density matrix of the isolated system in time. It
can be written as direct product of the forward, $U$, and backward, $U^\dag$,
system propagators,
\begin{align}
    K(S^\pm_n, S^\pm_{n+1}, \Delta t) &= U(S^+_n, S^+_{n+1}, \Delta t)\,U^\dag(S^-_n, S^-_{n+1}, \Delta t).
\end{align}

There are two parts to the simulation. First, we have to need a proper representation for the
forward-backward propagator. This is challenging due to the exponential growth
of space requirements with the number of particles in the system. Various
formalisms have been used to obtain the propagators in the compressed matrix
product operator form~\cite{paeckelTimeevolutionMethodsMatrixproduct2019,
schollwockDensitymatrixRenormalizationGroup2011a}. MS-TNPI can work with any of
these propagators, though we have used the second-order
Suzuki-Trotter split propagator.

\begin{figure}
    \includegraphics[scale=0.2]{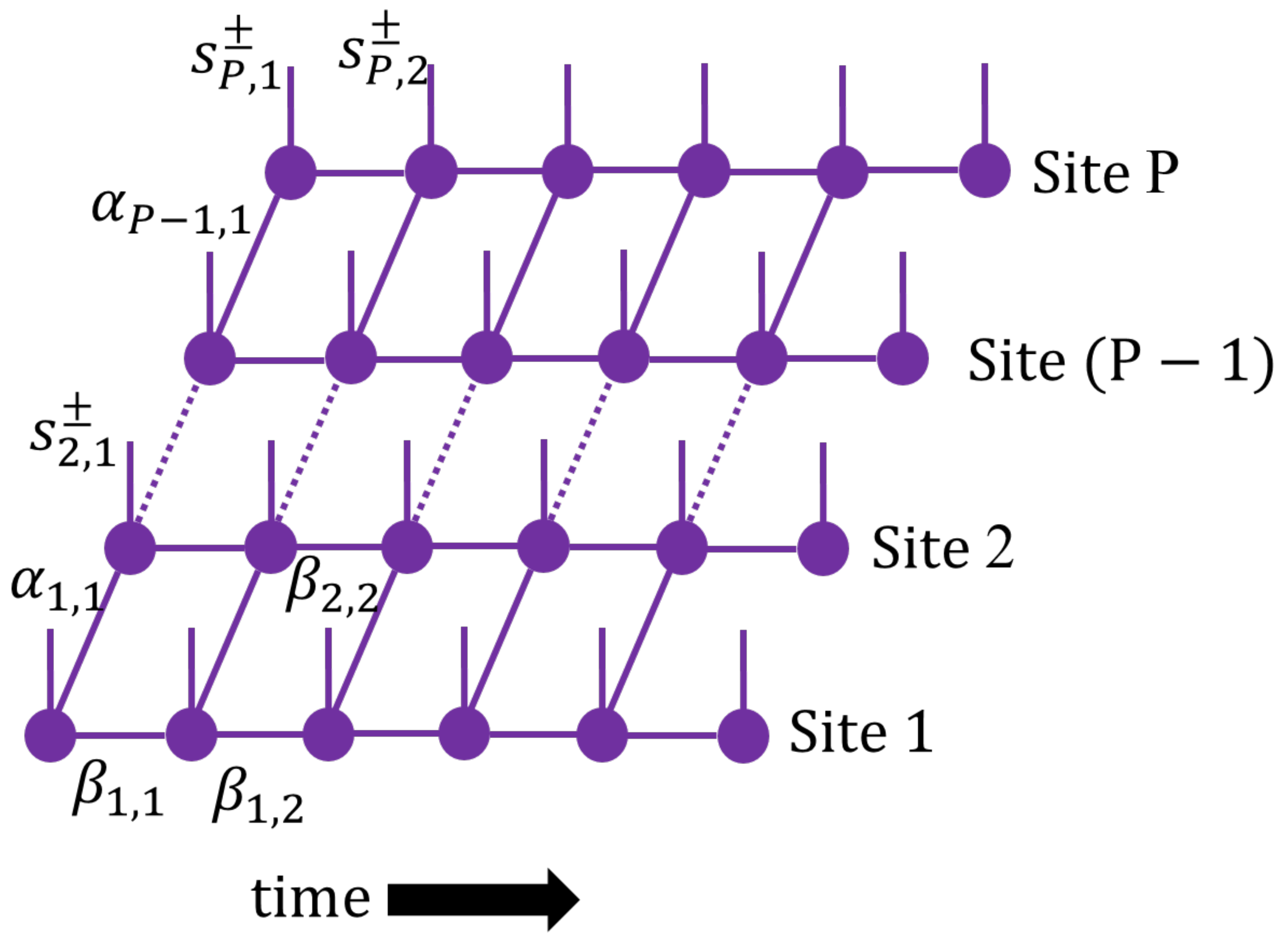}
    \caption{2D MS-TNPI tensor network.}\label{fig:ms-tnpi}
\end{figure}

Once the propagator has been defined, we need to include the influence
functional. For that, one needs to be able to account for the non-Markovian
memory induced by the bath. Usual propagations with MPO propagators simulate
the Markovian dynamics in absence of solvent
modes~\cite{whiteRealTimeEvolutionUsing2004,
yangTimedependentVariationalPrinciple2020}. To take the memory effects into
account, we need a compact tensor network representation of the bare path
amplitude tensor $P^{(0)}$. This intuitively involves the construction of a
grid of multiple points on the time axis. Combined with the MPO representation
for the forward-backward propagator that involves a splitting along the spatial
or system axis, one can visualize the formation of a 2D tensor network as shown
in Fig.~\ref{fig:ms-tnpi}. Once this network is created, it is possible to
apply the site-dependent influence functional, written as an MPO on every row
of the network. In this section, we derive the formalism required for
specifying and constructing the network. In Appendix~\ref{app:IF-MPO}, we deal with
representing the influence functional in the form of an MPO to be applied to
each row.

\begin{figure}
    \includegraphics[scale=0.3]{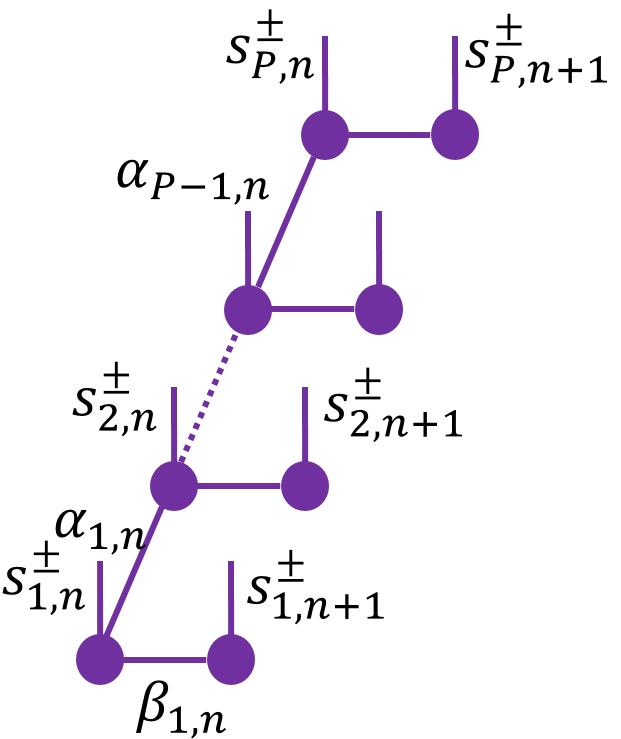}
    \caption{Schematic depiction of the SVD factorization of the forward-backward propagator MPO.}\label{fig:fbpropMPO}
\end{figure}

To obtain the tensor network schematically shown in Fig.~\ref{fig:ms-tnpi}, we
proceed by using SVD to factor the forward-backward propagator MPO as shown in Fig.~\ref{fig:fbpropMPO}:
\begin{align}
    K\left(S_{n}^\pm, S_{n+1}^\pm, \Delta t\right)  &=\sum_{\left\{\alpha_{(i,n)}\right\}}
    W^{s_{1,n}^\pm, s_{1,n+1}^\pm}_{\alpha_{(1,n)}}  W^{s_{2,n}^\pm,
    s_{2,n+1}^\pm}_{\alpha_{(1,n)},\alpha_{(2,n)} }\nonumber                      \\
                                                      & \cdots W^{s_{P-1,n}^\pm,
                                     s_{P-1,n+1}^\pm}_{\alpha_{(P-2,n)},\alpha_{(P-1,n)}}W^{s_{P,n}^\pm,
                             s_{P,n+1}^\pm}_{\alpha_{(P-1,n)}},\label{eq:prop_MPO}\\
  W^{s_{1,n}^\pm, s_{1,n+1}^\pm}_{\alpha_{(1,n)}}                   & = \sum_{\beta_{(1,n)}} L^{s_{1,n}^\pm}_{\alpha_{(1,n)}, \beta_{(1,n)}} R^{s_{1,n+1}^\pm}_{\beta_{(1,n)}}                           \label{eq:fbmpo1}        \\
  W^{s_{i,n}^\pm, s_{i,n+1}^\pm}_{\alpha_{(i-1,n)},\alpha_{(i,n)} } & = \sum_{\beta_{(i,n)}} L^{s_{i,n}^\pm}_{\alpha_{(i-1,n)},\alpha_{(i,n)},\beta_{(i,n)}} R^{s_{i,n+1}^\pm}_{\beta_{(i,n)}},\medspace 1<i<P  \label{eq:fbmpo2} \\
  W^{s_{P,n}^\pm, s_{P,n+1}^\pm}_{\alpha_{(P-1,n)}}                 & = \sum_{\beta_{(P,n)}} L^{s_{P,n}^\pm}_{\alpha_{(P-1,n)}, \beta_{(P,n)}} R^{s_{P,n+1}^\pm}_{\beta_{(P,n)}}.\label{eq:fbmpo3}
\end{align}
where $W$ are the tensors constituting the MPO representation. $L$ and $R$ are
the factors obtained through the SVD procedure with the square root of the
singular value matrix being absorbed into the factors. As per convention, the
bonds along the spatial and temporal dimensions are denoted by $\alpha$ and
$\beta$. Notice that the tensors $L$ and $R$ act on single system sites. By
substituting the factorization for the forward-backward propagator in
Eq.~\ref{eq:pi} and regrouping terms, one can obtain the following form: 
\begin{align}
    P^{(0)}_{S_{0}^{\pm}\cdots S_{N}^{\pm}} & = \sum_{\left\{\beta_{n}\right\}} \mathbb{T}^{S_{0}^{\pm}}_{\beta_{0}}
    \cdots \mathbb{T}^{S_{n}^{\pm}}_{\beta_{n-1}, \beta_{n}}
    \cdots\mathbb{T}_{\beta_{N-1}}^{S_{N}^{\pm}}.\label{eq:MS-PA_MPS}
\end{align}

It is now quite simple to list the tensors constituting the MPs,
$\mathbb{T}$, in Eq.~\ref{eq:MS-PA_MPS}, in terms of $L$ and $R$. In the most
general case, each of these constituent tensors, represented here by $M$,
possesses five indices: one site, $s^\pm_{i,n}$, and four bonds
($\alpha_{(i,n)}$, $\beta_{(i,n)}$, $\alpha_{(i-1,n)}$ and $\beta_{(i,n-1)}$),
where the values of $i$ and $n$ correspond to the location of the tensor in the
2D grid structure. The tensors constituting the edges of the network
(Fig.~\ref{fig:ms-tnpi}) obviously have a different topology. The tensors
corresponding to the initial time point, or equivalently the first column, are
given as:
\begin{align}
  M^{s^\pm_{1,0}}_{\alpha_{(1,0)}, \beta_{(1,0)}}                   & = L^{s^\pm_{1,0}}_{\alpha_{(1,0)}, \beta_{(1,0)}}                   \\
  M^{s^\pm_{i,0}}_{\alpha_{(i,0)}, \beta_{(i,0)}, \alpha_{(i-1,0)}} & = L^{s^\pm_{i,0}}_{\alpha_{(i-1,0)}, \alpha_{(i,0)}, \beta_{(i,0)}} \\
  M^{s^\pm_{P,0}}_{\alpha_{(P-1, 0)}, \beta_{(P,0)}}                & = L^{s^\pm_{P,0}}_{\alpha_{(P-1,0)}, \beta_{(P,0)}}.
\end{align}
Next, we list the expressions for the final point, last column:
\begin{align}
  M^{s^\pm_{1,N}}_{\beta_{(1,N-1)}} & = R^{s^\pm_{1,N}}_{\beta_{(1,N-1)}}\label{eq:lastcol1}  \\
  M^{s^\pm_{i,N}}_{\beta_{(i,N-1)}} & = R^{s^\pm_{i,N}}_{\beta_{(i,N-1)}}\label{eq:lastcol2}  \\
  M^{s^\pm_{P,N}}_{\beta_{(P,N-1)}} & = R^{s^\pm_{P,N}}_{\beta_{(P,N-1)}}\label{eq:lastcol3}.
\end{align}
Lastly, for an intermediate time point, $n$:
\begin{align}
  M^{s^\pm_{1,n}}_{\alpha_{(1,n)}, \beta_{(1,n)}, \beta_{(1,n-1)}}                   & = R^{s^\pm_{1,n}}_{\beta_{(1,n-1)}} L^{s^\pm_{1,n}}_{\alpha_{(1,n)}, \beta_{(1,n)}}                   \\
  M^{s^\pm_{i,n}}_{\alpha_{(i,n)}, \beta_{(i,n)}, \alpha_{(i-1,n)}, \beta_{(i,n-1)}} & = R^{s^\pm_{i,n}}_{\beta_{(i,n-1)}} L^{s^\pm_{i,n}}_{\alpha_{(i-1,n)}, \alpha_{(i,n)}, \beta_{(i,n)}} \\
  M^{s^\pm_{P,n}}_{\beta_{(P,n)}, \alpha_{(P-1,n)}, \beta_{(P,n-1)}}                 & = R^{s^\pm_{P,n}}_{\beta_{(P,n-1)}} L^{s^\pm_{P,n}}_{\alpha_{(P-1,n)}, \beta_{(P,n)}}.
\end{align}

These expressions give a complete description of the MS-TNPI tensor network.
The final step is the inclusion of the influence functional. The tensor network
is ready for application of an MPO encoding the influence functional on each of
the rows. The form of the influence functional MPO is given in
Appendix~\ref{app:IF-MPO}.

\section{Influence Functional MPO}\label{app:IF-MPO}

After the construction of the network, we need to be able to define the
influence functional MPOs that enable the systematic incorporation of the
impact of the environment on the dynamics of the system. Summarizing the
results discussed in depth in Ref.~\cite{boseTensorNetworkRepresentation2021},
the full site-local influence functional,
$F_{i}\left[\left\{s^{\pm}_{i,n}\right\}\right]$, can be factored and rewritten
as a product of terms corresponding to interactions with different end-times,
$F_{i,k}\left[\left\{s^\pm_{i,n}\right\}\right]$:
\begin{align}
  F_{i}\left[\left\{s^{\pm}_{i,n}\right\}\right] &= \prod_{0\le k\le N}F_{i,k}\left[\left\{s^{\pm}_{i,n}\right\}\right]
\end{align} 
with 
\begin{align}
  F_{i,k}\left[\left\{s^\pm_{i,n}\right\}\right] & = \exp\left(-\frac{1}{\hbar}\Delta s_{i,k}\sum_{0\le k'\le k} (\Re(\eta_{kk'}^{i})\Delta s_{i,k'}\right.\nonumber \\
                                             & \left.\vphantom{\sum_{0\le k\le N}}+2i\Im(\eta_{kk'}^{i})\bar{s}_{i,k'})\right).\label{eq:if_last_pt}
\end{align}

Grouping the forward-backward states of the extended system by unique values of
$\Delta s_{i,k}$ allows us to represent the the influence functional associated
with a particular site and end time point,
$F_{i,k}\left[\left\{s^\pm_{i,n}\right\}\right]$, as an MPO. To see this, let us
consider the case where there are $b$ unique values of $\Delta s_{i,k}$ indexed
by $\beta$. We note that for the subset of forward-backward paths where $\Delta s_{i,k}=
f_i(\beta)$, we can express the influence functional, Eq.~\ref{eq:if_last_pt}, as 
\begin{align}
F_{i,k}\left(\beta\right)= F_{i,k}^{0}\left(\beta\right) \otimes F_{i,k}^{1}\left(\beta\right) 
                                                   \cdots F_{i,k}^{k-1}\left(\beta\right) \otimes F_{i,k}^{k}\left(\beta\right)
                                                    \mathcal{P}^{s_{i,k}^{\pm}}_{ f_i(\beta)},\label{eq:if_prod_op}                                             
\end{align}
where $F_{i,k}^{k'}\left(\beta\right)=e^{-\tfrac{1}{\hbar}
f_i(\beta)\left(\Re\left(\eta_{kk'}^{i}\right)\mathbb{D}_{i,k'} +
2i\Im\left(\eta_{kk'}^{i}\right)\mathbb{S}_{i,k'}\right)}$ is an operator that
only acts on the $i$\textsuperscript{th} site and the $k'$\textsuperscript{th}
time point. In this notation $\mathcal{P}^{s_{i,k}^{\pm}}_{f_{i}(\beta)}$ is the
projection operator on to the space where $\Delta s_{i,k} = f_{i}(\beta)$;
additionally, $\mathbb{D}_{i,n}$ and $\mathbb{S}_{i,n}$ are diagonal matrices
that represent the difference and average position of the system in the
forward-backward basis, respectively. For each of the $b$ unique value of
$\Delta s_{i,k}$, the expression for the influence functional reduces to a
direct product of local operators; therefore,
$F_{i,k}\left[\left\{s^\pm_{i,n}\right\}\right]$ corresponds to a sum of direct
products. Hence we can express $F_{i,k}\left[\left\{s^\pm_{i,n}\right\}\right]$
as a MPO, $\mathbb{F}_{i,k}$, with a bond dimension of $b$:

\begin{align}
  \mathbb{F}_{i,k} &= \sum_{\left\{\beta_{(i,n)}\right\}} W^{s_{i,0}^{\pm}, s_{i,0}^{'\pm}}_{\beta_{(i,0)}}(\eta_{k0}^i)\cdots W^{s_{i,k'}^{\pm}, s_{i,k'}^{'\pm}}_{\beta_{(i,k'-1)}, \beta_{(i,k')}} (\eta_{kk'}^{i})\nonumber\\
                 &\times W^{s_{i,k'+1}^{\pm}, s_{i,k'+1}^{'\pm}}_{\beta_{(i,k')}, \beta_{(i,k'+1)}} (\eta_{k(k'+1)}^i)\cdots W^{s_{i,k}^{\pm}, s_{i,k}^{'\pm}}_{\beta_{(i,k-1)}} (\eta_{kk}^{i})\label{eq:IFMPO}
\end{align}
Here, $W$ are the various tensors constituting our influence functional MPO, and
are defined as:
\begin{widetext}
\begin{align}
  W^{s_{i,0}^{\pm}, s_{i,0}^{'\pm}}_{\beta_{(i,0)}}(\eta_{k0}^{i}) &= 
    \delta_{s_{i,0}^{\pm}, s_{i,0}^{'\pm}}
    \exp\left(-\tfrac{1}{\hbar}f_i(\beta_{(i,0)}) (\Re(\eta_{k0}^{i}) \Delta s_{i,0} + 2i\Im(\eta_{k0}^{i}) \bar{s}_{i,0})\right)\label{eq:IFMPO_0}\\
  W^{s_{i,k'}^{\pm}, s_{i,k'}^{'\pm}}_{\beta_{(i,k'-1)}, \beta_{(i,k')}} (\eta_{kk'}^{i})&= 
    \delta_{s_{i,k'}^{\pm}, s_{i,k'}^{'\pm}}\delta_{\beta_{(i,k'-1)}, \beta_{(i,k')}}
    \exp\left(-\tfrac{1}{\hbar}f_i(\beta_{(i,k'-1)}) (\Re(\eta_{kk'}^{i}) \Delta s_{i,k'} + 2i\Im(\eta_{kk'}^{i}) \bar{s}_{i,k'})\right)\label{eq:IFMPO_kp}\\
  W^{s_{i,k}^{\pm}, s_{i,k}^{'\pm}}_{\beta_{(i,k-1)}} (\eta_{kk}^{i})&= 
    \delta_{s_{i,k}^{\pm}, s_{i,k}^{'\pm}} \mathcal{P}^{s_{i,k}^{\pm}}_{f_i(\beta_{(i,k-1)})}
    \exp\left(-\tfrac{1}{\hbar}\Delta s_{i,k} (\Re(\eta_{kk}^{i}) \Delta s_{i,k} + 2i\Im(\eta_{kk}^{i}) \bar{s}_{i,k})\right)\label{eq:IFMPO_k}
\end{align}
\end{widetext}
It's worth noting that the primed forward-backward indices that appear in
Eqs.~\ref{eq:IFMPO}\,--\,\ref{eq:IFMPO_k} are necessary for bookkeeping
purposes only. Computationally, for any particular time-step, we only apply the
$\mathbb{F}_{i,n}$ operators corresponding to the final time point. This
procedure leads to a sequential or iterative build-up of the full influence
functional including the effects arising from all the intermediate points.

\section{Total Dipole Moment MPO}\label{app:Dipole_MPO}

In the paper, we have outlined a general idea about how to calculate the MPO
representation for the total dipole operator as a sum of individual MPOs.
However, it is possible to represent it as an MPO in an exact manner without
resorting to MPO summations. In this appendix, we outline the basic formulae
involved in deriving a low bond-dimensioned MPO representation for the
operator.

Consider the initial ``state'' involved in the absorption spectrum:
\begin{align}
    C(t) &= \Tr(\hat{\mu}(t)\hat{\mu}(0)\rho(0)).
\end{align}
The $\hat\mu$ operator acts on the forward space or on the ket side. We can
formulate an MPO which is a delta function on the bra side and the total dipole
operator on the ket side. This MPO is denoted by $\bar{\bar\mu}^+$. The
superscript ``+'' denotes that the total dipole acts on the forward space.

\begin{align}
    \bar{\bar\mu}^+ &= \sum_{\left\{\alpha_{i}\right\}} W^{s_{1}^{\pm}, s_{1}^{'\pm}}_{\alpha_{1}} W^{s_{2}^{\pm}, s_{2}^{'\pm}}_{\alpha_{1},\alpha_{2} } 
    \cdots W^{s_{P-1}^{\pm}, s_{P-1}^{'\pm}}_{\alpha_{P-2},\alpha_{P-1}} W^{s_{P}^{\pm}, s_{P}^{'\pm}}_{\alpha_{P-1}}\label{eq:Dipole_MPO}
\end{align}
where $W$ are the constituent tensors of the MPO. The site indices are
$s_i^\pm$ and $s_i^{\pm'}$. The intuition is that $s_i^\pm$ are the ``input''
indices and $s_i^{'\pm}$ are the output indices. The primes have no other
semantic meaning. The bond index connecting the $i$\textsuperscript{th} and
$(i+1)$\textsuperscript{th} are denoted by $\alpha_i$.

Below we list the explicit formulae for the $W$ tensors.
\begin{align}
    W^{s_{1}^{\pm}, s_{1}^{'\pm}}_{\alpha_{1}} &= 
    \begin{cases}
       \mel{s^{'+}_1}{\hat{\mu}_1}{s^{+}_1}  \delta_{s_{1}^{-}, s_{1}^{'-}}, &\text{$\alpha_{1}=0$}\\
        \delta_{s_{1}^{\pm}, s_{1}^{'\pm}},  &\text{$\alpha_{1}=1$}
    \end{cases}\\
    W^{s_{i}^{\pm}, s_{i}^{'\pm}}_{\alpha_{i-1},\alpha_{i}} &= 
    \begin{cases}
        \delta_{s_{i}^{\pm}, s_{i}^{'\pm}},  &\text{$\alpha_{i-1}=\alpha_{i}$}\\
        \mel{s^{'+}_i}{\hat{\mu}_i}{s^{+}_i} \delta_{s_{i}^{-}, s_{i}^{'-}} , &\text{$\alpha_{i-1}=1$ and $\alpha_{i}=0$}\\
        0, &\text{otherwise}
    \end{cases}\\
    W^{s_{P}^{\pm}, s_{P}^{'\pm}}_{\alpha_{P-1}} &= 
    \begin{cases}
        \delta_{s_{P}^{\pm}, s_{P}^{'\pm}},  &\text{$\alpha_{P-1}=1$}\\
        \mel{s^{'+}_P}{\hat{\mu}_P}{s^{+}_P} \delta_{s_{P}^{-}, s_{P}^{'-}} , &\text{$\alpha_{P-1}=0$}
    \end{cases}
\end{align}
Note that since the bond indices only take two values (0 or 1), the bond
dimension of this analytical dipole moment MPO is exactly 2. 

\section{Excitonic Eigenstates}\label{app:exciton}
In the main text, we discussed three different system parameter sets in the
context of the absorption spectrum. We dealt with the TrEsp parameters, with
and without the \SI{197}{\per\cm} variation in the excitation energies of
neighboring BChl units, and the experimentally derived parameters with the
variation. Here we list the energies of the excitonic eigenstates for all three
parameters. These eigen-energies do not account for the solvent interaction at
all.

\begin{table}[H]
    \centering
    \rowcolors{3}{white}{gray!25}
    \begin{tabular}{c|c|c|c}
        \rowcolor{gray!50} 
        State & TrEsp & TrEsp & Experimental\\
        \rowcolor{gray!35} 
        No. & w/o Variation & w/ Variation & w/ Variation\\\hline
        0 & 11679.5 & 11664.4 & 11715.4 \\
        $\pm 1$ & 11703.1 & 11686.8 & 11756.6 \\
        $\pm 2$ & 11770.0 & 11749.1 & 11872.9 \\
        $\pm 3$ & 11868.9 & 11834.5 & 12039.4 \\
        4 & 11959.5 & 11888.6 & 12163.1 \\
        5 & 12025.5 & 12096.4 & 12404.8 \\
        $\pm 6$ & 12116.1 & 12150.6 & 12528.5 \\
        $\pm 7$ & 12215.1 & 12235.9 & 12695.0 \\
        $\pm 8$ & 12282.0 & 12298.3 & 12811.3 \\
        9 & 12305.5 & 12320.7 & 12852.6 \\
    \end{tabular}
    \caption{Energies for the excitonic eigenstates in absence of coupling to the thermal environment.}
\end{table}

\bibliography{library}

\begin{thebibliography}{54}%
\makeatletter
\providecommand \@ifxundefined [1]{%
 \@ifx{#1\undefined}
}%
\providecommand \@ifnum [1]{%
 \ifnum #1\expandafter \@firstoftwo
 \else \expandafter \@secondoftwo
 \fi
}%
\providecommand \@ifx [1]{%
 \ifx #1\expandafter \@firstoftwo
 \else \expandafter \@secondoftwo
 \fi
}%
\providecommand \natexlab [1]{#1}%
\providecommand \enquote  [1]{``#1''}%
\providecommand \bibnamefont  [1]{#1}%
\providecommand \bibfnamefont [1]{#1}%
\providecommand \citenamefont [1]{#1}%
\providecommand \href@noop [0]{\@secondoftwo}%
\providecommand \href [0]{\begingroup \@sanitize@url \@href}%
\providecommand \@href[1]{\@@startlink{#1}\@@href}%
\providecommand \@@href[1]{\endgroup#1\@@endlink}%
\providecommand \@sanitize@url [0]{\catcode `\\12\catcode `\$12\catcode
  `\&12\catcode `\#12\catcode `\^12\catcode `\_12\catcode `\%12\relax}%
\providecommand \@@startlink[1]{}%
\providecommand \@@endlink[0]{}%
\providecommand \url  [0]{\begingroup\@sanitize@url \@url }%
\providecommand \@url [1]{\endgroup\@href {#1}{\urlprefix }}%
\providecommand \urlprefix  [0]{URL }%
\providecommand \Eprint [0]{\href }%
\providecommand \doibase [0]{https://doi.org/}%
\providecommand \selectlanguage [0]{\@gobble}%
\providecommand \bibinfo  [0]{\@secondoftwo}%
\providecommand \bibfield  [0]{\@secondoftwo}%
\providecommand \translation [1]{[#1]}%
\providecommand \BibitemOpen [0]{}%
\providecommand \bibitemStop [0]{}%
\providecommand \bibitemNoStop [0]{.\EOS\space}%
\providecommand \EOS [0]{\spacefactor3000\relax}%
\providecommand \BibitemShut  [1]{\csname bibitem#1\endcsname}%
\let\auto@bib@innerbib\@empty
%</preamble>
\bibitem [{\citenamefont {Ishizaki}\ and\ \citenamefont
  {Fleming}(2009{\natexlab{a}})}]{ishizakiAdequacyRedfieldEquation2009}%
  \BibitemOpen
  \bibfield  {author} {\bibinfo {author} {\bibfnamefont {A.}~\bibnamefont
  {Ishizaki}}\ and\ \bibinfo {author} {\bibfnamefont {G.~R.}\ \bibnamefont
  {Fleming}},\ }\bibfield  {title} {\bibinfo {title} {On the adequacy of the
  {{Redfield}} equation and related approaches to the study of quantum dynamics
  in electronic energy transfer},\ }\href {https://doi.org/10.1063/1.3155214}
  {\bibfield  {journal} {\bibinfo  {journal} {J. Chem. Phys.}\ }\textbf
  {\bibinfo {volume} {130}},\ \bibinfo {pages} {234110} (\bibinfo {year}
  {2009}{\natexlab{a}})}\BibitemShut {NoStop}%
\bibitem [{\citenamefont {Novoderezhkin}\ and\ \citenamefont
  {Razjivin}(1996)}]{novoderezhkinTheTheoryofForster1996}%
  \BibitemOpen
  \bibfield  {author} {\bibinfo {author} {\bibfnamefont {V.}~\bibnamefont
  {Novoderezhkin}}\ and\ \bibinfo {author} {\bibfnamefont {A.}~\bibnamefont
  {Razjivin}},\ }\bibfield  {title} {\bibinfo {title} {The theory of
  forster-type migration between clusters of strongly interacting molecules:
  application to light-harvesting complexes of purple bacteria},\ }\href
  {https://doi.org/https://doi.org/10.1016/0301-0104(96)00130-9} {\bibfield
  {journal} {\bibinfo  {journal} {Chemical Physics}\ }\textbf {\bibinfo
  {volume} {211}},\ \bibinfo {pages} {203} (\bibinfo {year}
  {1996})}\BibitemShut {NoStop}%
\bibitem [{\citenamefont {White}(1992)}]{whiteDensityMatrixFormulation1992}%
  \BibitemOpen
  \bibfield  {author} {\bibinfo {author} {\bibfnamefont {S.~R.}\ \bibnamefont
  {White}},\ }\bibfield  {title} {\bibinfo {title} {Density matrix formulation
  for quantum renormalization groups},\ }\href
  {https://doi.org/10.1103/physrevlett.69.2863} {\bibfield  {journal} {\bibinfo
   {journal} {Phys. Rev. Lett.}\ }\textbf {\bibinfo {volume} {69}},\ \bibinfo
  {pages} {2863} (\bibinfo {year} {1992})}\BibitemShut {NoStop}%
\bibitem [{\citenamefont {White}\ and\ \citenamefont
  {Feiguin}(2004)}]{whiteRealTimeEvolutionUsing2004}%
  \BibitemOpen
  \bibfield  {author} {\bibinfo {author} {\bibfnamefont {S.~R.}\ \bibnamefont
  {White}}\ and\ \bibinfo {author} {\bibfnamefont {A.~E.}\ \bibnamefont
  {Feiguin}},\ }\bibfield  {title} {\bibinfo {title} {Real-{{Time Evolution
  Using}} the {{Density Matrix Renormalization Group}}},\ }\bibfield  {journal}
  {\bibinfo  {journal} {Phys. Rev. Lett.}\ }\textbf {\bibinfo {volume} {93}},\
  \href {https://doi.org/10.1103/physrevlett.93.076401}
  {10.1103/physrevlett.93.076401} (\bibinfo {year} {2004})\BibitemShut
  {NoStop}%
\bibitem [{\citenamefont
  {Schollw{\"o}ck}(2005)}]{schollwockDensitymatrixRenormalizationGroup2005}%
  \BibitemOpen
  \bibfield  {author} {\bibinfo {author} {\bibfnamefont {U.}~\bibnamefont
  {Schollw{\"o}ck}},\ }\bibfield  {title} {\bibinfo {title} {The density-matrix
  renormalization group},\ }\href {https://doi.org/10.1103/revmodphys.77.259}
  {\bibfield  {journal} {\bibinfo  {journal} {Rev. Mod. Phys.}\ }\textbf
  {\bibinfo {volume} {77}},\ \bibinfo {pages} {259} (\bibinfo {year}
  {2005})}\BibitemShut {NoStop}%
\bibitem [{\citenamefont
  {Schollw{\"o}ck}(2011{\natexlab{a}})}]{schollwockDensitymatrixRenormalizationGroup2011a}%
  \BibitemOpen
  \bibfield  {author} {\bibinfo {author} {\bibfnamefont {U.}~\bibnamefont
  {Schollw{\"o}ck}},\ }\bibfield  {title} {\bibinfo {title} {The density-matrix
  renormalization group in the age of matrix product states},\ }\href
  {https://doi.org/10.1016/j.aop.2010.09.012} {\bibfield  {journal} {\bibinfo
  {journal} {Ann. Phys. (N. Y.)}\ }\textbf {\bibinfo {volume} {326}},\ \bibinfo
  {pages} {96} (\bibinfo {year} {2011}{\natexlab{a}})}\BibitemShut {NoStop}%
\bibitem [{\citenamefont {Jiang}\ \emph {et~al.}(2020)\citenamefont {Jiang},
  \citenamefont {Li}, \citenamefont {Ren},\ and\ \citenamefont
  {Shuai}}]{jiangFiniteTemperatureDynamical2020}%
  \BibitemOpen
  \bibfield  {author} {\bibinfo {author} {\bibfnamefont {T.}~\bibnamefont
  {Jiang}}, \bibinfo {author} {\bibfnamefont {W.}~\bibnamefont {Li}}, \bibinfo
  {author} {\bibfnamefont {J.}~\bibnamefont {Ren}},\ and\ \bibinfo {author}
  {\bibfnamefont {Z.}~\bibnamefont {Shuai}},\ }\bibfield  {title} {\bibinfo
  {title} {{Finite Temperature Dynamical Density Matrix Renormalization Group
  for Spectroscopy in Frequency Domain}},\ }\href
  {https://doi.org/10.1021/acs.jpclett.0c00905} {\bibfield  {journal} {\bibinfo
   {journal} {J. Phys. Chem. Lett.}\ }\textbf {\bibinfo {volume} {11}},\
  \bibinfo {pages} {3761} (\bibinfo {year} {2020})}\BibitemShut {NoStop}%
\bibitem [{\citenamefont
  {Beck}(2000)}]{beckMulticonfigurationTimedependentHartree2000}%
  \BibitemOpen
  \bibfield  {author} {\bibinfo {author} {\bibfnamefont {M.}~\bibnamefont
  {Beck}},\ }\bibfield  {title} {\bibinfo {title} {The multiconfiguration
  time-dependent {{Hartree}} ({{MCTDH}}) method: A highly efficient algorithm
  for propagating wavepackets},\ }\href
  {https://doi.org/10.1016/s0370-1573(99)00047-2} {\bibfield  {journal}
  {\bibinfo  {journal} {Physics Reports}\ }\textbf {\bibinfo {volume} {324}},\
  \bibinfo {pages} {1} (\bibinfo {year} {2000})}\BibitemShut {NoStop}%
\bibitem [{\citenamefont {Wang}\ and\ \citenamefont
  {Thoss}(2003)}]{wangMultilayerFormulationMulticonfiguration2003}%
  \BibitemOpen
  \bibfield  {author} {\bibinfo {author} {\bibfnamefont {H.}~\bibnamefont
  {Wang}}\ and\ \bibinfo {author} {\bibfnamefont {M.}~\bibnamefont {Thoss}},\
  }\bibfield  {title} {\bibinfo {title} {Multilayer formulation of the
  multiconfiguration time-dependent {{Hartree}} theory},\ }\href
  {https://doi.org/10.1063/1.1580111} {\bibfield  {journal} {\bibinfo
  {journal} {J. Chem. Phys.}\ }\textbf {\bibinfo {volume} {119}},\ \bibinfo
  {pages} {1289} (\bibinfo {year} {2003})}\BibitemShut {NoStop}%
\bibitem [{\citenamefont {Ren}\ \emph {et~al.}(2018)\citenamefont {Ren},
  \citenamefont {Shuai},\ and\ \citenamefont {{Kin-Lic
  Chan}}}]{renTimeDependentDensityMatrix2018}%
  \BibitemOpen
  \bibfield  {author} {\bibinfo {author} {\bibfnamefont {J.}~\bibnamefont
  {Ren}}, \bibinfo {author} {\bibfnamefont {Z.}~\bibnamefont {Shuai}},\ and\
  \bibinfo {author} {\bibfnamefont {G.}~\bibnamefont {{Kin-Lic Chan}}},\
  }\bibfield  {title} {\bibinfo {title} {Time-{{Dependent Density Matrix
  Renormalization Group Algorithms}} for {{Nearly Exact Absorption}} and
  {{Fluorescence Spectra}} of {{Molecular Aggregates}} at {{Both Zero}} and
  {{Finite Temperature}}},\ }\href {https://doi.org/10.1021/acs.jctc.8b00628}
  {\bibfield  {journal} {\bibinfo  {journal} {J. Chem. Theory Comput.}\
  }\textbf {\bibinfo {volume} {14}},\ \bibinfo {pages} {5027} (\bibinfo {year}
  {2018})}\BibitemShut {NoStop}%
\bibitem [{\citenamefont
  {Tanimura}(2020)}]{tanimuraNumericallyExactApproach2020}%
  \BibitemOpen
  \bibfield  {author} {\bibinfo {author} {\bibfnamefont {Y.}~\bibnamefont
  {Tanimura}},\ }\bibfield  {title} {\bibinfo {title} {Numerically ``exact''
  approach to open quantum dynamics: The hierarchical equations of motion
  ({{HEOM}})},\ }\href {https://doi.org/10.1063/5.0011599} {\bibfield
  {journal} {\bibinfo  {journal} {J. Chem. Phys.}\ }\textbf {\bibinfo {volume}
  {153}},\ \bibinfo {pages} {020901} (\bibinfo {year} {2020})}\BibitemShut
  {NoStop}%
\bibitem [{\citenamefont {Tanimura}\ and\ \citenamefont
  {Kubo}(1989)}]{tanimuraTimeEvolutionQuantum1989}%
  \BibitemOpen
  \bibfield  {author} {\bibinfo {author} {\bibfnamefont {Y.}~\bibnamefont
  {Tanimura}}\ and\ \bibinfo {author} {\bibfnamefont {R.}~\bibnamefont
  {Kubo}},\ }\bibfield  {title} {\bibinfo {title} {Time {{Evolution}} of a
  {{Quantum System}} in {{Contact}} with a {{Nearly Gaussian}}-{{Markoffian
  Noise Bath}}},\ }\href {https://doi.org/10.1143/JPSJ.58.101} {\bibfield
  {journal} {\bibinfo  {journal} {J. Phys. Soc. Jpn.}\ }\textbf {\bibinfo
  {volume} {58}},\ \bibinfo {pages} {101} (\bibinfo {year} {1989})}\BibitemShut
  {NoStop}%
\bibitem [{\citenamefont {Yan}\ \emph {et~al.}(2021)\citenamefont {Yan},
  \citenamefont {Xu}, \citenamefont {Li},\ and\ \citenamefont
  {Shi}}]{yanEfficientPropagationoftheHierarchical2021}%
  \BibitemOpen
  \bibfield  {author} {\bibinfo {author} {\bibfnamefont {Y.}~\bibnamefont
  {Yan}}, \bibinfo {author} {\bibfnamefont {M.}~\bibnamefont {Xu}}, \bibinfo
  {author} {\bibfnamefont {T.}~\bibnamefont {Li}},\ and\ \bibinfo {author}
  {\bibfnamefont {Q.}~\bibnamefont {Shi}},\ }\bibfield  {title} {\bibinfo
  {title} {{Efficient propagation of the hierarchical equations of motion using
  the Tucker and hierarchical Tucker tensors}},\ }\href
  {https://doi.org/10.1063/5.0050720} {\bibfield  {journal} {\bibinfo
  {journal} {J. Chem. Phys.}\ }\textbf {\bibinfo {volume} {154}},\ \bibinfo
  {pages} {194104} (\bibinfo {year} {2021})}\BibitemShut {NoStop}%
\bibitem [{\citenamefont {Makri}\ and\ \citenamefont
  {Makarov}(1995{\natexlab{a}})}]{makriTensorPropagatorIterative1995}%
  \BibitemOpen
  \bibfield  {author} {\bibinfo {author} {\bibfnamefont {N.}~\bibnamefont
  {Makri}}\ and\ \bibinfo {author} {\bibfnamefont {D.~E.}\ \bibnamefont
  {Makarov}},\ }\bibfield  {title} {\bibinfo {title} {Tensor propagator for
  iterative quantum time evolution of reduced density matrices. {{I}}.
  {{Theory}}},\ }\href {https://doi.org/10.1063/1.469508} {\bibfield  {journal}
  {\bibinfo  {journal} {J. Chem. Phys.}\ }\textbf {\bibinfo {volume} {102}},\
  \bibinfo {pages} {4600} (\bibinfo {year} {1995}{\natexlab{a}})}\BibitemShut
  {NoStop}%
\bibitem [{\citenamefont {Makri}\ and\ \citenamefont
  {Makarov}(1995{\natexlab{b}})}]{makriTensorPropagatorIterative1995a}%
  \BibitemOpen
  \bibfield  {author} {\bibinfo {author} {\bibfnamefont {N.}~\bibnamefont
  {Makri}}\ and\ \bibinfo {author} {\bibfnamefont {D.~E.}\ \bibnamefont
  {Makarov}},\ }\bibfield  {title} {\bibinfo {title} {Tensor propagator for
  iterative quantum time evolution of reduced density matrices. {{II}}.
  {{Numerical}} methodology},\ }\href {https://doi.org/10.1063/1.469509}
  {\bibfield  {journal} {\bibinfo  {journal} {J. Chem. Phys.}\ }\textbf
  {\bibinfo {volume} {102}},\ \bibinfo {pages} {4611} (\bibinfo {year}
  {1995}{\natexlab{b}})}\BibitemShut {NoStop}%
\bibitem [{\citenamefont {Str{\"{u}}mpfer}\ and\ \citenamefont
  {Schulten}(2009)}]{strumpferLightHarvestingComplexII2009}%
  \BibitemOpen
  \bibfield  {author} {\bibinfo {author} {\bibfnamefont {J.}~\bibnamefont
  {Str{\"{u}}mpfer}}\ and\ \bibinfo {author} {\bibfnamefont {K.}~\bibnamefont
  {Schulten}},\ }\bibfield  {title} {\bibinfo {title} {{Light harvesting
  complex II B850 excitation dynamics}},\ }\href
  {https://doi.org/10.1063/1.3271348} {\bibfield  {journal} {\bibinfo
  {journal} {J. Chem. Phys.}\ }\textbf {\bibinfo {volume} {131}},\ \bibinfo
  {pages} {225101} (\bibinfo {year} {2009})}\BibitemShut {NoStop}%
\bibitem [{\citenamefont {Str{\"u}mpfer}\ and\ \citenamefont
  {Schulten}(2011)}]{strumpferEffectCorrelatedBath2011}%
  \BibitemOpen
  \bibfield  {author} {\bibinfo {author} {\bibfnamefont {J.}~\bibnamefont
  {Str{\"u}mpfer}}\ and\ \bibinfo {author} {\bibfnamefont {K.}~\bibnamefont
  {Schulten}},\ }\bibfield  {title} {\bibinfo {title} {The effect of correlated
  bath fluctuations on exciton transfer},\ }\href
  {https://doi.org/10.1063/1.3557042} {\bibfield  {journal} {\bibinfo
  {journal} {J. Chem. Phys.}\ }\textbf {\bibinfo {volume} {134}},\ \bibinfo
  {pages} {095102} (\bibinfo {year} {2011})}\BibitemShut {NoStop}%
\bibitem [{\citenamefont {Str{\"u}mpfer}\ and\ \citenamefont
  {Schulten}(2012)}]{strumpferExcitedStateDynamics2012}%
  \BibitemOpen
  \bibfield  {author} {\bibinfo {author} {\bibfnamefont {J.}~\bibnamefont
  {Str{\"u}mpfer}}\ and\ \bibinfo {author} {\bibfnamefont {K.}~\bibnamefont
  {Schulten}},\ }\bibfield  {title} {\bibinfo {title} {Excited state dynamics
  in photosynthetic reaction center and light harvesting complex 1},\ }\href
  {https://doi.org/10.1063/1.4738953} {\bibfield  {journal} {\bibinfo
  {journal} {J. Chem. Phys.}\ }\textbf {\bibinfo {volume} {137}},\ \bibinfo
  {pages} {065101} (\bibinfo {year} {2012})}\BibitemShut {NoStop}%
\bibitem [{\citenamefont {Makri}(2018)}]{makriModularPathIntegral2018}%
  \BibitemOpen
  \bibfield  {author} {\bibinfo {author} {\bibfnamefont {N.}~\bibnamefont
  {Makri}},\ }\bibfield  {title} {\bibinfo {title} {Modular path integral
  methodology for real-time quantum dynamics},\ }\href
  {https://doi.org/10.1063/1.5058223} {\bibfield  {journal} {\bibinfo
  {journal} {J. Chem. Phys.}\ }\textbf {\bibinfo {volume} {149}},\ \bibinfo
  {pages} {214108} (\bibinfo {year} {2018})}\BibitemShut {NoStop}%
\bibitem [{\citenamefont {Kundu}\ and\ \citenamefont
  {Makri}(2020)}]{kunduRealTimePathIntegral2020}%
  \BibitemOpen
  \bibfield  {author} {\bibinfo {author} {\bibfnamefont {S.}~\bibnamefont
  {Kundu}}\ and\ \bibinfo {author} {\bibfnamefont {N.}~\bibnamefont {Makri}},\
  }\bibfield  {title} {\bibinfo {title} {Real-{{Time Path Integral Simulation}}
  of {{Exciton}}-{{Vibration Dynamics}} in {{Light}}-{{Harvesting
  Bacteriochlorophyll Aggregates}}},\ }\href
  {https://doi.org/10.1021/acs.jpclett.0c02760} {\bibfield  {journal} {\bibinfo
   {journal} {J. Phys. Chem. Lett.}\ }\textbf {\bibinfo {volume} {11}},\
  \bibinfo {pages} {8783} (\bibinfo {year} {2020})}\BibitemShut {NoStop}%
\bibitem [{\citenamefont {Kundu}\ and\ \citenamefont
  {Makri}(2021)}]{kunduOriginVibrationalFeatures2021}%
  \BibitemOpen
  \bibfield  {author} {\bibinfo {author} {\bibfnamefont {S.}~\bibnamefont
  {Kundu}}\ and\ \bibinfo {author} {\bibfnamefont {N.}~\bibnamefont {Makri}},\
  }\bibfield  {title} {\bibinfo {title} {Origin of vibrational features in the
  excitation energy transfer dynamics of perylene bisimide {{J}}-aggregates},\
  }\href {https://doi.org/10.1063/5.0041514} {\bibfield  {journal} {\bibinfo
  {journal} {J. Chem. Phys.}\ }\textbf {\bibinfo {volume} {154}},\ \bibinfo
  {pages} {114301} (\bibinfo {year} {2021})}\BibitemShut {NoStop}%
\bibitem [{\citenamefont {Bose}\ and\ \citenamefont
  {Walters}(2022)}]{boseMultisiteDecompositionTensor2022}%
  \BibitemOpen
  \bibfield  {author} {\bibinfo {author} {\bibfnamefont {A.}~\bibnamefont
  {Bose}}\ and\ \bibinfo {author} {\bibfnamefont {P.~L.}\ \bibnamefont
  {Walters}},\ }\bibfield  {title} {\bibinfo {title} {{A multisite
  decomposition of the tensor network path integrals}},\ }\href
  {https://doi.org/10.1063/5.0073234} {\bibfield  {journal} {\bibinfo
  {journal} {J. Chem. Phys.}\ }\textbf {\bibinfo {volume} {156}},\ \bibinfo
  {pages} {24101} (\bibinfo {year} {2022})}\BibitemShut {NoStop}%
\bibitem [{\citenamefont {Lee}\ \emph {et~al.}(2016)\citenamefont {Lee},
  \citenamefont {Huo},\ and\ \citenamefont
  {Coker}}]{leeSemiclassicalPathIntegral2016}%
  \BibitemOpen
  \bibfield  {author} {\bibinfo {author} {\bibfnamefont {M.~K.}\ \bibnamefont
  {Lee}}, \bibinfo {author} {\bibfnamefont {P.}~\bibnamefont {Huo}},\ and\
  \bibinfo {author} {\bibfnamefont {D.~F.}\ \bibnamefont {Coker}},\ }\bibfield
  {title} {\bibinfo {title} {Semiclassical {{Path Integral Dynamics}}:
  Photosynthetic {{Energy Transfer}} with {{Realistic Environment
  Interactions}}},\ }\href
  {https://doi.org/10.1146/annurev-physchem-040215-112252} {\bibfield
  {journal} {\bibinfo  {journal} {Annu. Rev. Phys. Chem.}\ }\textbf {\bibinfo
  {volume} {67}},\ \bibinfo {pages} {639} (\bibinfo {year} {2016})}\BibitemShut
  {NoStop}%
\bibitem [{\citenamefont {Bose}\ and\ \citenamefont
  {Walters}(2021)}]{boseTensorNetworkRepresentation2021}%
  \BibitemOpen
  \bibfield  {author} {\bibinfo {author} {\bibfnamefont {A.}~\bibnamefont
  {Bose}}\ and\ \bibinfo {author} {\bibfnamefont {P.~L.}\ \bibnamefont
  {Walters}},\ }\bibfield  {title} {\bibinfo {title} {A tensor network
  representation of path integrals: Implementation and analysis},\ }\href@noop
  {} {\bibfield  {journal} {\bibinfo  {journal} {arXiv pre-print server}\ }
  (\bibinfo {year} {2021})}\BibitemShut {NoStop}%
\bibitem [{\citenamefont {Bose}(2022)}]{bosePairwiseConnectedTensor2022}%
  \BibitemOpen
  \bibfield  {author} {\bibinfo {author} {\bibfnamefont {A.}~\bibnamefont
  {Bose}},\ }\bibfield  {title} {\bibinfo {title} {{Pairwise connected tensor
  network representation of path integrals}},\ }\href
  {https://doi.org/10.1103/PhysRevB.105.024309} {\bibfield  {journal} {\bibinfo
   {journal} {Phys. Rev. B}\ }\textbf {\bibinfo {volume} {105}},\ \bibinfo
  {pages} {024309} (\bibinfo {year} {2022})}\BibitemShut {NoStop}%
\bibitem [{\citenamefont {Strathearn}\ \emph {et~al.}(2018)\citenamefont
  {Strathearn}, \citenamefont {Kirton}, \citenamefont {Kilda}, \citenamefont
  {Keeling},\ and\ \citenamefont
  {Lovett}}]{strathearnEfficientNonMarkovianQuantum2018}%
  \BibitemOpen
  \bibfield  {author} {\bibinfo {author} {\bibfnamefont {A.}~\bibnamefont
  {Strathearn}}, \bibinfo {author} {\bibfnamefont {P.}~\bibnamefont {Kirton}},
  \bibinfo {author} {\bibfnamefont {D.}~\bibnamefont {Kilda}}, \bibinfo
  {author} {\bibfnamefont {J.}~\bibnamefont {Keeling}},\ and\ \bibinfo {author}
  {\bibfnamefont {B.~W.}\ \bibnamefont {Lovett}},\ }\bibfield  {title}
  {\bibinfo {title} {Efficient non-{{Markovian}} quantum dynamics using
  time-evolving matrix product operators},\ }\bibfield  {journal} {\bibinfo
  {journal} {Nat. Commun}\ }\textbf {\bibinfo {volume} {9}},\ \href
  {https://doi.org/10.1038/s41467-018-05617-3} {10.1038/s41467-018-05617-3}
  (\bibinfo {year} {2018})\BibitemShut {NoStop}%
\bibitem [{\citenamefont {J{\o}rgensen}\ and\ \citenamefont
  {Pollock}(2019)}]{jorgensenExploitingCausalTensor2019}%
  \BibitemOpen
  \bibfield  {author} {\bibinfo {author} {\bibfnamefont {M.~R.}\ \bibnamefont
  {J{\o}rgensen}}\ and\ \bibinfo {author} {\bibfnamefont {F.~A.}\ \bibnamefont
  {Pollock}},\ }\bibfield  {title} {\bibinfo {title} {Exploiting the {{Causal
  Tensor Network Structure}} of {{Quantum Processes}} to {{Efficiently Simulate
  Non}}-{{Markovian Path Integrals}}},\ }\bibfield  {journal} {\bibinfo
  {journal} {Phys. Rev. Lett.}\ }\textbf {\bibinfo {volume} {123}},\ \href
  {https://doi.org/10.1103/physrevlett.123.240602}
  {10.1103/physrevlett.123.240602} (\bibinfo {year} {2019})\BibitemShut
  {NoStop}%
\bibitem [{\citenamefont
  {Schollw{\"o}ck}(2011{\natexlab{b}})}]{schollwockDensitymatrixRenormalizationGroup2011}%
  \BibitemOpen
  \bibfield  {author} {\bibinfo {author} {\bibfnamefont {U.}~\bibnamefont
  {Schollw{\"o}ck}},\ }\bibfield  {title} {\bibinfo {title} {The density-matrix
  renormalization group: A short introduction},\ }\href
  {https://doi.org/10.1098/rsta.2010.0382} {\bibfield  {journal} {\bibinfo
  {journal} {Philos. Trans. A Math. Phys. Eng. Sci.}\ }\textbf {\bibinfo
  {volume} {369}},\ \bibinfo {pages} {2643} (\bibinfo {year}
  {2011}{\natexlab{b}})}\BibitemShut {NoStop}%
\bibitem [{\citenamefont {Paeckel}\ \emph {et~al.}(2019)\citenamefont
  {Paeckel}, \citenamefont {K{\"o}hler}, \citenamefont {Swoboda}, \citenamefont
  {Manmana}, \citenamefont {Schollw{\"o}ck},\ and\ \citenamefont
  {Hubig}}]{paeckelTimeevolutionMethodsMatrixproduct2019}%
  \BibitemOpen
  \bibfield  {author} {\bibinfo {author} {\bibfnamefont {S.}~\bibnamefont
  {Paeckel}}, \bibinfo {author} {\bibfnamefont {T.}~\bibnamefont {K{\"o}hler}},
  \bibinfo {author} {\bibfnamefont {A.}~\bibnamefont {Swoboda}}, \bibinfo
  {author} {\bibfnamefont {S.~R.}\ \bibnamefont {Manmana}}, \bibinfo {author}
  {\bibfnamefont {U.}~\bibnamefont {Schollw{\"o}ck}},\ and\ \bibinfo {author}
  {\bibfnamefont {C.}~\bibnamefont {Hubig}},\ }\bibfield  {title} {\bibinfo
  {title} {Time-evolution methods for matrix-product states},\ }\href
  {https://doi.org/10.1016/j.aop.2019.167998} {\bibfield  {journal} {\bibinfo
  {journal} {Ann. Phys. (N. Y.)}\ }\textbf {\bibinfo {volume} {411}},\ \bibinfo
  {pages} {167998} (\bibinfo {year} {2019})}\BibitemShut {NoStop}%
\bibitem [{\citenamefont {Feynman}\ and\ \citenamefont
  {Vernon}(1963)}]{feynmanTheoryGeneralQuantum1963}%
  \BibitemOpen
  \bibfield  {author} {\bibinfo {author} {\bibfnamefont {R.~P.}\ \bibnamefont
  {Feynman}}\ and\ \bibinfo {author} {\bibfnamefont {F.~L.}\ \bibnamefont
  {Vernon}},\ }\bibfield  {title} {\bibinfo {title} {The theory of a general
  quantum system interacting with a linear dissipative system},\ }\href
  {https://doi.org/10.1016/0003-4916(63)90068-x} {\bibfield  {journal}
  {\bibinfo  {journal} {Ann. Phys. (N. Y.)}\ }\textbf {\bibinfo {volume}
  {24}},\ \bibinfo {pages} {118} (\bibinfo {year} {1963})}\BibitemShut
  {NoStop}%
\bibitem [{\citenamefont {Ishizaki}\ and\ \citenamefont
  {Fleming}(2009{\natexlab{b}})}]{ishizakiTheoreticalExaminationQuantum2009}%
  \BibitemOpen
  \bibfield  {author} {\bibinfo {author} {\bibfnamefont {A.}~\bibnamefont
  {Ishizaki}}\ and\ \bibinfo {author} {\bibfnamefont {G.~R.}\ \bibnamefont
  {Fleming}},\ }\bibfield  {title} {\bibinfo {title} {Theoretical examination
  of quantum coherence in a photosynthetic system at physiological
  temperature},\ }\href {https://doi.org/10.1073/pnas.0908989106} {\bibfield
  {journal} {\bibinfo  {journal} {Proc. Natl. Acad. Sci.}\ }\textbf {\bibinfo
  {volume} {106}},\ \bibinfo {pages} {17255} (\bibinfo {year}
  {2009}{\natexlab{b}})}\BibitemShut {NoStop}%
\bibitem [{\citenamefont {R{\"a}tsep}\ and\ \citenamefont
  {Freiberg}(2007)}]{ratsepElectronPhononVibronic2007}%
  \BibitemOpen
  \bibfield  {author} {\bibinfo {author} {\bibfnamefont {M.}~\bibnamefont
  {R{\"a}tsep}}\ and\ \bibinfo {author} {\bibfnamefont {A.}~\bibnamefont
  {Freiberg}},\ }\bibfield  {title} {\bibinfo {title} {Electron\textendash
  phonon and vibronic couplings in the {{FMO}} bacteriochlorophyll a antenna
  complex studied by difference fluorescence line narrowing},\ }\href
  {https://doi.org/10.1016/j.jlumin.2007.02.053} {\bibfield  {journal}
  {\bibinfo  {journal} {J. Lumin.}\ }\textbf {\bibinfo {volume} {127}},\
  \bibinfo {pages} {251} (\bibinfo {year} {2007})}\BibitemShut {NoStop}%
\bibitem [{\citenamefont {Bose}\ and\ \citenamefont
  {Makri}(2020)}]{boseAllModeQuantumClassical2020}%
  \BibitemOpen
  \bibfield  {author} {\bibinfo {author} {\bibfnamefont {A.}~\bibnamefont
  {Bose}}\ and\ \bibinfo {author} {\bibfnamefont {N.}~\bibnamefont {Makri}},\
  }\bibfield  {title} {\bibinfo {title} {All-{{Mode
  Quantum}}\textendash{{Classical Path Integral Simulation}} of
  {{Bacteriochlorophyll Dimer Exciton}}-{{Vibration Dynamics}}},\ }\href
  {https://doi.org/10.1021/acs.jpcb.0c03032} {\bibfield  {journal} {\bibinfo
  {journal} {J. Phys. Chem. B}\ }\textbf {\bibinfo {volume} {124}},\ \bibinfo
  {pages} {5028} (\bibinfo {year} {2020})}\BibitemShut {NoStop}%
\bibitem [{\citenamefont {Olbrich}\ and\ \citenamefont
  {Kleinekath{\"o}fer}(2010)}]{olbrichTimeDependentAtomisticView2010}%
  \BibitemOpen
  \bibfield  {author} {\bibinfo {author} {\bibfnamefont {C.}~\bibnamefont
  {Olbrich}}\ and\ \bibinfo {author} {\bibfnamefont {U.}~\bibnamefont
  {Kleinekath{\"o}fer}},\ }\bibfield  {title} {\bibinfo {title}
  {Time-{{Dependent Atomistic View}} on the {{Electronic Relaxation}} in
  {{Light}}-{{Harvesting System II}}},\ }\href
  {https://doi.org/10.1021/jp106542v} {\bibfield  {journal} {\bibinfo
  {journal} {J. Phys. Chem. B}\ }\textbf {\bibinfo {volume} {114}},\ \bibinfo
  {pages} {12427} (\bibinfo {year} {2010})}\BibitemShut {NoStop}%
\bibitem [{\citenamefont {Olbrich}\ \emph {et~al.}(2011)\citenamefont
  {Olbrich}, \citenamefont {Str{\"u}mpfer}, \citenamefont {Schulten},\ and\
  \citenamefont
  {Kleinekath{\"o}fer}}]{olbrichTheorySimulationEnvironmental2011}%
  \BibitemOpen
  \bibfield  {author} {\bibinfo {author} {\bibfnamefont {C.}~\bibnamefont
  {Olbrich}}, \bibinfo {author} {\bibfnamefont {J.}~\bibnamefont
  {Str{\"u}mpfer}}, \bibinfo {author} {\bibfnamefont {K.}~\bibnamefont
  {Schulten}},\ and\ \bibinfo {author} {\bibfnamefont {U.}~\bibnamefont
  {Kleinekath{\"o}fer}},\ }\bibfield  {title} {\bibinfo {title} {Theory and
  {{Simulation}} of the {{Environmental Effects}} on {{FMO Electronic
  Transitions}}},\ }\href {https://doi.org/10.1021/jz2007676} {\bibfield
  {journal} {\bibinfo  {journal} {J. Phys. Chem. Lett.}\ }\textbf {\bibinfo
  {volume} {2}},\ \bibinfo {pages} {1771} (\bibinfo {year} {2011})}\BibitemShut
  {NoStop}%
\bibitem [{\citenamefont {Maity}\ \emph {et~al.}(2020)\citenamefont {Maity},
  \citenamefont {Bold}, \citenamefont {Prajapati}, \citenamefont {Sokolov},
  \citenamefont {Kuba{\v r}}, \citenamefont {Elstner},\ and\ \citenamefont
  {Kleinekath{\"o}fer}}]{maityDFTBMMMolecular2020}%
  \BibitemOpen
  \bibfield  {author} {\bibinfo {author} {\bibfnamefont {S.}~\bibnamefont
  {Maity}}, \bibinfo {author} {\bibfnamefont {B.~M.}\ \bibnamefont {Bold}},
  \bibinfo {author} {\bibfnamefont {J.~D.}\ \bibnamefont {Prajapati}}, \bibinfo
  {author} {\bibfnamefont {M.}~\bibnamefont {Sokolov}}, \bibinfo {author}
  {\bibfnamefont {T.}~\bibnamefont {Kuba{\v r}}}, \bibinfo {author}
  {\bibfnamefont {M.}~\bibnamefont {Elstner}},\ and\ \bibinfo {author}
  {\bibfnamefont {U.}~\bibnamefont {Kleinekath{\"o}fer}},\ }\bibfield  {title}
  {\bibinfo {title} {{{DFTB}}/{{MM Molecular Dynamics Simulations}} of the
  {{FMO Light}}-{{Harvesting Complex}}},\ }\href
  {https://doi.org/10.1021/acs.jpclett.0c02526} {\bibfield  {journal} {\bibinfo
   {journal} {J. Phys. Chem. Lett.}\ }\textbf {\bibinfo {volume} {11}},\
  \bibinfo {pages} {8660} (\bibinfo {year} {2020})}\BibitemShut {NoStop}%
\bibitem [{\citenamefont {Caldeira}\ and\ \citenamefont
  {Leggett}(1983)}]{caldeiraPathIntegralApproach1983}%
  \BibitemOpen
  \bibfield  {author} {\bibinfo {author} {\bibfnamefont {A.~O.}\ \bibnamefont
  {Caldeira}}\ and\ \bibinfo {author} {\bibfnamefont {A.~J.}\ \bibnamefont
  {Leggett}},\ }\bibfield  {title} {\bibinfo {title} {Path integral approach to
  quantum {{Brownian}} motion},\ }\href
  {https://doi.org/10.1016/0378-4371(83)90013-4} {\bibfield  {journal}
  {\bibinfo  {journal} {Physica A: Statistical Mechanics and its Applications}\
  }\textbf {\bibinfo {volume} {121}},\ \bibinfo {pages} {587} (\bibinfo {year}
  {1983})}\BibitemShut {NoStop}%
\bibitem [{\citenamefont {Makri}(1999)}]{makriLinearResponseApproximation1999}%
  \BibitemOpen
  \bibfield  {author} {\bibinfo {author} {\bibfnamefont {N.}~\bibnamefont
  {Makri}},\ }\bibfield  {title} {\bibinfo {title} {The {{Linear Response
  Approximation}} and {{Its Lowest Order Corrections}}:\, {{An Influence
  Functional Approach}}},\ }\href {https://doi.org/10.1021/jp9847540}
  {\bibfield  {journal} {\bibinfo  {journal} {J. Phys. Chem. B}\ }\textbf
  {\bibinfo {volume} {103}},\ \bibinfo {pages} {2823} (\bibinfo {year}
  {1999})}\BibitemShut {NoStop}%
\bibitem [{\citenamefont {Kim}\ and\ \citenamefont
  {Rossky}(2002)}]{kimEvaluationQuantumCorrelation2002}%
  \BibitemOpen
  \bibfield  {author} {\bibinfo {author} {\bibfnamefont {H.}~\bibnamefont
  {Kim}}\ and\ \bibinfo {author} {\bibfnamefont {P.~J.}\ \bibnamefont
  {Rossky}},\ }\bibfield  {title} {\bibinfo {title} {Evaluation of {{Quantum
  Correlation Functions}} from {{Classical Data}}},\ }\href
  {https://doi.org/10.1021/jp020669n} {\bibfield  {journal} {\bibinfo
  {journal} {J. Phys. Chem. B}\ }\textbf {\bibinfo {volume} {106}},\ \bibinfo
  {pages} {8240} (\bibinfo {year} {2002})}\BibitemShut {NoStop}%
\bibitem [{\citenamefont {Kim}\ and\ \citenamefont
  {Rossky}(2006)}]{kimEvaluationQuantumCorrelation2006}%
  \BibitemOpen
  \bibfield  {author} {\bibinfo {author} {\bibfnamefont {H.}~\bibnamefont
  {Kim}}\ and\ \bibinfo {author} {\bibfnamefont {P.~J.}\ \bibnamefont
  {Rossky}},\ }\bibfield  {title} {\bibinfo {title} {Evaluation of quantum
  correlation functions from classical data: {{Anharmonic}} models},\ }\href
  {https://doi.org/10.1063/1.2274412} {\bibfield  {journal} {\bibinfo
  {journal} {The Journal of Chemical Physics}\ }\textbf {\bibinfo {volume}
  {125}},\ \bibinfo {pages} {074107} (\bibinfo {year} {2006})},\ \Eprint
  {https://arxiv.org/abs/https://doi.org/10.1063/1.2274412}
  {https://doi.org/10.1063/1.2274412} \BibitemShut {NoStop}%
\bibitem [{\citenamefont {Valleau}\ \emph {et~al.}(2012)\citenamefont
  {Valleau}, \citenamefont {Eisfeld},\ and\ \citenamefont
  {Aspuru-Guzik}}]{valleauOnTheAlternatives2012}%
  \BibitemOpen
  \bibfield  {author} {\bibinfo {author} {\bibfnamefont {S.}~\bibnamefont
  {Valleau}}, \bibinfo {author} {\bibfnamefont {A.}~\bibnamefont {Eisfeld}},\
  and\ \bibinfo {author} {\bibfnamefont {A.}~\bibnamefont {Aspuru-Guzik}},\
  }\bibfield  {title} {\bibinfo {title} {{On the alternatives for bath
  correlators and spectral densities from mixed quantum-classical
  simulations}},\ }\href {https://doi.org/10.1063/1.4769079} {\bibfield
  {journal} {\bibinfo  {journal} {J. Chem. Phys.}\ }\textbf {\bibinfo {volume}
  {137}},\ \bibinfo {pages} {224103} (\bibinfo {year} {2012})}\BibitemShut
  {NoStop}%
\bibitem [{\citenamefont {Lee}\ and\ \citenamefont
  {Coker}(2016)}]{leeModelingElectronicNuclearInteractions2016}%
  \BibitemOpen
  \bibfield  {author} {\bibinfo {author} {\bibfnamefont {M.~K.}\ \bibnamefont
  {Lee}}\ and\ \bibinfo {author} {\bibfnamefont {D.~F.}\ \bibnamefont
  {Coker}},\ }\bibfield  {title} {\bibinfo {title} {Modeling
  {{Electronic}}-{{Nuclear Interactions}} for {{Excitation Energy Transfer
  Processes}} in {{Light}}-{{Harvesting Complexes}}},\ }\href
  {https://doi.org/10.1021/acs.jpclett.6b01440} {\bibfield  {journal} {\bibinfo
   {journal} {J. Phys. Chem. Lett.}\ }\textbf {\bibinfo {volume} {7}},\
  \bibinfo {pages} {3171} (\bibinfo {year} {2016})}\BibitemShut {NoStop}%
\bibitem [{\citenamefont {Daley}\ \emph {et~al.}(2004)\citenamefont {Daley},
  \citenamefont {Kollath}, \citenamefont {Schollw{\"o}ck},\ and\ \citenamefont
  {Vidal}}]{daleyTimedependentDensitymatrixRenormalizationgroup2004}%
  \BibitemOpen
  \bibfield  {author} {\bibinfo {author} {\bibfnamefont {A.~J.}\ \bibnamefont
  {Daley}}, \bibinfo {author} {\bibfnamefont {C.}~\bibnamefont {Kollath}},
  \bibinfo {author} {\bibfnamefont {U.}~\bibnamefont {Schollw{\"o}ck}},\ and\
  \bibinfo {author} {\bibfnamefont {G.}~\bibnamefont {Vidal}},\ }\bibfield
  {title} {\bibinfo {title} {Time-dependent density-matrix
  renormalization-group using adaptive effective {{Hilbert}} spaces},\ }\href
  {https://doi.org/10.1088/1742-5468/2004/04/p04005} {\bibfield  {journal}
  {\bibinfo  {journal} {J. Stat. Mech. Theory Exp.}\ }\textbf {\bibinfo
  {volume} {2004}},\ \bibinfo {pages} {P04005} (\bibinfo {year}
  {2004})}\BibitemShut {NoStop}%
\bibitem [{\citenamefont {Haegeman}\ \emph {et~al.}(2011)\citenamefont
  {Haegeman}, \citenamefont {Cirac}, \citenamefont {Osborne}, \citenamefont
  {Pi{\v z}orn}, \citenamefont {Verschelde},\ and\ \citenamefont
  {Verstraete}}]{haegemanTimeDependentVariationalPrinciple2011}%
  \BibitemOpen
  \bibfield  {author} {\bibinfo {author} {\bibfnamefont {J.}~\bibnamefont
  {Haegeman}}, \bibinfo {author} {\bibfnamefont {J.~I.}\ \bibnamefont {Cirac}},
  \bibinfo {author} {\bibfnamefont {T.~J.}\ \bibnamefont {Osborne}}, \bibinfo
  {author} {\bibfnamefont {I.}~\bibnamefont {Pi{\v z}orn}}, \bibinfo {author}
  {\bibfnamefont {H.}~\bibnamefont {Verschelde}},\ and\ \bibinfo {author}
  {\bibfnamefont {F.}~\bibnamefont {Verstraete}},\ }\bibfield  {title}
  {\bibinfo {title} {Time-{{Dependent Variational Principle}} for {{Quantum
  Lattices}}},\ }\href {https://doi.org/10.1103/PhysRevLett.107.070601}
  {\bibfield  {journal} {\bibinfo  {journal} {Phys. Rev. Lett.}\ }\textbf
  {\bibinfo {volume} {107}},\ \bibinfo {pages} {070601} (\bibinfo {year}
  {2011})}\BibitemShut {NoStop}%
\bibitem [{\citenamefont {Yang}\ and\ \citenamefont
  {White}(2020)}]{yangTimedependentVariationalPrinciple2020}%
  \BibitemOpen
  \bibfield  {author} {\bibinfo {author} {\bibfnamefont {M.}~\bibnamefont
  {Yang}}\ and\ \bibinfo {author} {\bibfnamefont {S.~R.}\ \bibnamefont
  {White}},\ }\bibfield  {title} {\bibinfo {title} {Time-dependent variational
  principle with ancillary {{Krylov}} subspace},\ }\href
  {https://doi.org/10.1103/PhysRevB.102.094315} {\bibfield  {journal} {\bibinfo
   {journal} {Phys. Rev. B}\ }\textbf {\bibinfo {volume} {102}},\ \bibinfo
  {pages} {094315} (\bibinfo {year} {2020})}\BibitemShut {NoStop}%
\bibitem [{\citenamefont {Zaletel}\ \emph {et~al.}(2015)\citenamefont
  {Zaletel}, \citenamefont {Mong}, \citenamefont {Karrasch}, \citenamefont
  {Moore},\ and\ \citenamefont {Pollmann}}]{Zaletel2015a}%
  \BibitemOpen
  \bibfield  {author} {\bibinfo {author} {\bibfnamefont {M.~P.}\ \bibnamefont
  {Zaletel}}, \bibinfo {author} {\bibfnamefont {R.~S.~K.}\ \bibnamefont
  {Mong}}, \bibinfo {author} {\bibfnamefont {C.}~\bibnamefont {Karrasch}},
  \bibinfo {author} {\bibfnamefont {J.~E.}\ \bibnamefont {Moore}},\ and\
  \bibinfo {author} {\bibfnamefont {F.}~\bibnamefont {Pollmann}},\ }\bibfield
  {title} {\bibinfo {title} {{Time-evolving a matrix product state with
  long-ranged interactions}},\ }\href
  {https://doi.org/10.1103/PhysRevB.91.165112} {\bibfield  {journal} {\bibinfo
  {journal} {Phys. Rev. B}\ }\textbf {\bibinfo {volume} {91}},\ \bibinfo
  {pages} {165112} (\bibinfo {year} {2015})}\BibitemShut {NoStop}%
\bibitem [{\citenamefont
  {Vidal}(2004)}]{vidalEfficientSimulationOneDimensional2004}%
  \BibitemOpen
  \bibfield  {author} {\bibinfo {author} {\bibfnamefont {G.}~\bibnamefont
  {Vidal}},\ }\bibfield  {title} {\bibinfo {title} {{Efficient Simulation of
  One-Dimensional Quantum Many-Body Systems}},\ }\href
  {https://doi.org/10.1103/PhysRevLett.93.040502} {\bibfield  {journal}
  {\bibinfo  {journal} {Phys. Rev. Lett.}\ }\textbf {\bibinfo {volume} {93}},\
  \bibinfo {pages} {040502} (\bibinfo {year} {2004})}\BibitemShut {NoStop}%
\bibitem [{\citenamefont
  {Renger}(2009)}]{rengerTheoryExcitationEnergyTransfer2009}%
  \BibitemOpen
  \bibfield  {author} {\bibinfo {author} {\bibfnamefont {T.}~\bibnamefont
  {Renger}},\ }\bibfield  {title} {\bibinfo {title} {{Theory of excitation
  energy transfer: from structure to function}},\ }\href
  {https://doi.org/10.1007/s11120-009-9472-9} {\bibfield  {journal} {\bibinfo
  {journal} {Photosynth. Res.}\ }\textbf {\bibinfo {volume} {102}},\ \bibinfo
  {pages} {471} (\bibinfo {year} {2009})}\BibitemShut {NoStop}%
\bibitem [{\citenamefont {Madjet}\ \emph {et~al.}(2006)\citenamefont {Madjet},
  \citenamefont {Abdurahman},\ and\ \citenamefont
  {Renger}}]{madjetIntermolecularCoulombCouplings2006}%
  \BibitemOpen
  \bibfield  {author} {\bibinfo {author} {\bibfnamefont {M.~E.}\ \bibnamefont
  {Madjet}}, \bibinfo {author} {\bibfnamefont {A.}~\bibnamefont {Abdurahman}},\
  and\ \bibinfo {author} {\bibfnamefont {T.}~\bibnamefont {Renger}},\
  }\bibfield  {title} {\bibinfo {title} {{Intermolecular Coulomb Couplings from
  Ab Initio Electrostatic Potentials: Application to Optical Transitions of
  Strongly Coupled Pigments in Photosynthetic Antennae and Reaction Centers}},\
  }\href {https://doi.org/10.1021/jp0615398} {\bibfield  {journal} {\bibinfo
  {journal} {J. Phys. Chem. B}\ }\textbf {\bibinfo {volume} {110}},\ \bibinfo
  {pages} {17268} (\bibinfo {year} {2006})}\BibitemShut {NoStop}%
\bibitem [{\citenamefont {Freiberg}\ \emph {et~al.}(2009)\citenamefont
  {Freiberg}, \citenamefont {Rätsep}, \citenamefont {Timpmann},\ and\
  \citenamefont {Trinkunas}}]{freibergExcitedStateDynamics2009}%
  \BibitemOpen
  \bibfield  {author} {\bibinfo {author} {\bibfnamefont {A.}~\bibnamefont
  {Freiberg}}, \bibinfo {author} {\bibfnamefont {M.}~\bibnamefont {Rätsep}},
  \bibinfo {author} {\bibfnamefont {K.}~\bibnamefont {Timpmann}},\ and\
  \bibinfo {author} {\bibfnamefont {G.}~\bibnamefont {Trinkunas}},\ }\bibfield
  {title} {\bibinfo {title} {Excitonic polarons in quasi-one-dimensional lh1
  and lh2 bacteriochlorophyll a antenna aggregates from photosynthetic
  bacteria: A wavelength-dependent selective spectroscopy study},\ }\href
  {https://doi.org/https://doi.org/10.1016/j.chemphys.2008.10.043} {\bibfield
  {journal} {\bibinfo  {journal} {Chemical Physics}\ }\textbf {\bibinfo
  {volume} {357}},\ \bibinfo {pages} {102} (\bibinfo {year} {2009})},\ \bibinfo
  {note} {excited State Dynamics in Light Harvesting Materials}\BibitemShut
  {NoStop}%
\bibitem [{\citenamefont {Tretiak}\ \emph {et~al.}(2000)\citenamefont
  {Tretiak}, \citenamefont {Middleton}, \citenamefont {Chernyak},\ and\
  \citenamefont {Mukamel}}]{tretiakBacteriochlorophyllCarotenoid2000}%
  \BibitemOpen
  \bibfield  {author} {\bibinfo {author} {\bibfnamefont {S.}~\bibnamefont
  {Tretiak}}, \bibinfo {author} {\bibfnamefont {C.}~\bibnamefont {Middleton}},
  \bibinfo {author} {\bibfnamefont {V.}~\bibnamefont {Chernyak}},\ and\
  \bibinfo {author} {\bibfnamefont {S.}~\bibnamefont {Mukamel}},\ }\bibfield
  {title} {\bibinfo {title} {{Bacteriochlorophyll and Carotenoid Excitonic
  Couplings in the LH2 System of Purple Bacteria}},\ }\href
  {https://doi.org/10.1021/jp001585m} {\bibfield  {journal} {\bibinfo
  {journal} {J. Phys. Chem. B}\ }\textbf {\bibinfo {volume} {104}},\ \bibinfo
  {pages} {9540} (\bibinfo {year} {2000})}\BibitemShut {NoStop}%
\bibitem [{\citenamefont {Hu}\ \emph {et~al.}(1997)\citenamefont {Hu},
  \citenamefont {Ritz}, \citenamefont {Damjanovi{\'{c}}},\ and\ \citenamefont
  {Schulten}}]{Hu1997a}%
  \BibitemOpen
  \bibfield  {author} {\bibinfo {author} {\bibfnamefont {X.}~\bibnamefont
  {Hu}}, \bibinfo {author} {\bibfnamefont {T.}~\bibnamefont {Ritz}}, \bibinfo
  {author} {\bibfnamefont {A.}~\bibnamefont {Damjanovi{\'{c}}}},\ and\ \bibinfo
  {author} {\bibfnamefont {K.}~\bibnamefont {Schulten}},\ }\bibfield  {title}
  {\bibinfo {title} {{Pigment Organization and Transfer of Electronic
  Excitation in the Photosynthetic Unit of Purple Bacteria}},\ }\href
  {https://doi.org/10.1021/jp963777g} {\bibfield  {journal} {\bibinfo
  {journal} {J. Phys. Chem. B}\ }\textbf {\bibinfo {volume} {101}},\ \bibinfo
  {pages} {3854} (\bibinfo {year} {1997})}\BibitemShut {NoStop}%
\bibitem [{\citenamefont {Damjanovi{\'{c}}i}\ \emph {et~al.}(2002)\citenamefont
  {Damjanovi{\'{c}}i}, \citenamefont {Kosztin}, \citenamefont
  {Kleinekath{\"{o}}fer},\ and\ \citenamefont
  {Schulten}}]{damjanoviciExcitonsinPhotosynthetic2002}%
  \BibitemOpen
  \bibfield  {author} {\bibinfo {author} {\bibfnamefont {A.}~\bibnamefont
  {Damjanovi{\'{c}}i}}, \bibinfo {author} {\bibfnamefont {I.}~\bibnamefont
  {Kosztin}}, \bibinfo {author} {\bibfnamefont {U.}~\bibnamefont
  {Kleinekath{\"{o}}fer}},\ and\ \bibinfo {author} {\bibfnamefont
  {K.}~\bibnamefont {Schulten}},\ }\bibfield  {title} {\bibinfo {title}
  {{Excitons in a photosynthetic light-harvesting system: A combined molecular
  dynamics, quantum chemistry, and polaron model study}},\ }\href
  {https://doi.org/10.1103/PhysRevE.65.031919} {\bibfield  {journal} {\bibinfo
  {journal} {Phys. Rev. E}\ }\textbf {\bibinfo {volume} {65}},\ \bibinfo
  {pages} {31919} (\bibinfo {year} {2002})}\BibitemShut {NoStop}%
\bibitem [{\citenamefont {Chen}\ \emph {et~al.}(2009)\citenamefont {Chen},
  \citenamefont {Zheng}, \citenamefont {Shi},\ and\ \citenamefont
  {Yan}}]{chenOpticalLineShapes2009}%
  \BibitemOpen
  \bibfield  {author} {\bibinfo {author} {\bibfnamefont {L.}~\bibnamefont
  {Chen}}, \bibinfo {author} {\bibfnamefont {R.}~\bibnamefont {Zheng}},
  \bibinfo {author} {\bibfnamefont {Q.}~\bibnamefont {Shi}},\ and\ \bibinfo
  {author} {\bibfnamefont {Y.}~\bibnamefont {Yan}},\ }\bibfield  {title}
  {\bibinfo {title} {{Optical line shapes of molecular aggregates: Hierarchical
  equations of motion method}},\ }\href {https://doi.org/10.1063/1.3213013}
  {\bibfield  {journal} {\bibinfo  {journal} {J. Chem. Phys.}\ }\textbf
  {\bibinfo {volume} {131}},\ \bibinfo {pages} {94502} (\bibinfo {year}
  {2009})}\BibitemShut {NoStop}%
\end{thebibliography}%
\end{document}